\documentclass[numberedappendix,twocolappendix]{emulateapj} 

\usepackage{epsfig,color,pifont}

\usepackage{natbib}
\bibpunct{(}{)}{;}{a}{}{,}
\def\spose#1{\hbox to 0pt{#1\hss}}
\def\ltsimm{\mathrel{\spose{\lower 3pt\hbox{$\sim$}}
        \raise 2.0pt\hbox{$<$}}}
\def\gtsimm{\mathrel{\spose{\lower 3pt\hbox{$\sim$}}
        \raise 2.0pt\hbox{$>$}}}

\def\km{{\rm\thinspace km}}
\def\cm{{\rm\thinspace cm}}

\def\s{{\rm\thinspace s}}
\def\yr{{\rm\thinspace yr}}
\def\g{{\rm\thinspace g}}
\def\kmps{\hbox{${\rm\km\s^{-1}\,}$}}
\def\cmps{\hbox{${\rm\cm\s^{-1}\,}$}}
\def\erg{{\rm\thinspace erg}}
\def\Hz{{\rm\thinspace Hz}}

\def\ster{{\rm\thinspace ster}}
\def\ergps{\hbox{${\rm\erg\s^{-1}\,}$}}
\def\Rsol{\hbox{${\rm\thinspace R_{\odot}}$}}
\def\Msol{\hbox{${\rm\thinspace M_{\odot}}$}}

\def\Msolpyr{\hbox{${\rm\Msol\yr^{-1}\,}$}}
\def\pcm{\hbox{${\rm\cm^{-1}\,}$}}
\def\pcm2{\hbox{${\rm\cm^{-2}\,}$}}
\def\pcm3{\hbox{${\rm\cm^{-3}\,}$}}
\def\ergpscm3Hz{\hbox{${\rm\ergps\cm^{-3}\Hz^{-1}\,}$}}
\def\ergpscm3Hzster{\hbox{${\rm\ergps\cm^{-3}\Hz^{-1}\ster^{-1}\,}$}}
\def\gpcm3{\hbox{${\rm\g\cm^{-3}\,}$}}
\def\ergpcm2{\hbox{${\rm\erg\cm^{-2}\,}$}}
\def\ergpcm3{\hbox{${\rm\erg\cm^{-3}\,}$}}
\def\phpscm2{\hbox{${\rm photons\s^{-1}\cm^{-2}\,}$}}

\def\Lx{\hbox{$L_{\rm X}$}}
\def\aap{{\rm A\&A}}
\def\apj{{\rm ApJ}}
\def\apjs{{\rm ApJS}}
\def\aj{{\rm AJ}}
\def\mnras{{\rm MNRAS}}

\begin{document} 

\title{Self regulated shocks in massive star binary systems}

\shorttitle{Self regulated shocks in massive binaries}

\author{E.~R.~Parkin$^{1}$ \& S.~A.~Sim$^{1,2}$\\
  $^{1}$Research School of Astronomy and Astrophysics, The Australian
  National University, Australia \\$^{2}$Astrophysics Research Centre,
  School of Mathematics and Physics, Queen's University Belfast,
  Belfast BT7 1NN}

\email{email: parkin@mso.anu.edu.au, s.sim@qub.ac.uk}

\shortauthors{E.~R.~Parkin \& S.~A.~Sim}
\label{firstpage}

\begin{abstract}
  In an early-type, massive star binary system, X-ray bright shocks
  result from the powerful collision of stellar winds driven by
  radiation pressure on spectral line transitions. We examine the
  influence of the X-rays from the wind-wind collision shocks on the
  radiative driving of the stellar winds using steady state models
  that include a parameterized line force with X-ray ionization
  dependence. Our primary result is that X-ray radiation from the
  shocks inhibits wind acceleration and can lead to a lower pre-shock
  velocity, and a correspondingly lower shocked plasma temperature,
  yet the {\it intrinsic} X-ray luminosity of the shocks, $L_{\rm X}$
  remains largely unaltered, with the exception of a modest increase
  at small binary separations. Due to the feedback loop between the
  ionizing X-rays from the shocks and the wind-driving, we term this
  scenario as {\it self regulated shocks}. This effect is found to
  greatly increase the range of binary separations at which a
  wind-photosphere collision is likely to occur in systems where the
  momenta of the two winds are significantly different. Furthermore,
  the excessive levels of X-ray ionization close to the shocks
  completely suppresses the line force, and we suggest that this may
  render radiative braking less effective. Comparisons of model
  results against observations reveals reasonable agreement in terms
  of $\log(L_{\rm X}/L_{\rm bol})$. The inclusion of self regulated
  shocks improves the match for $kT$ values in roughly equal wind
  momenta systems, but there is a systematic offset for systems with
  unequal wind momenta (if considered to be a wind-photosphere
  collision).
\end{abstract}


\maketitle

\section{Introduction}
\label{sec:intro}

A large fraction of massive stars reside in binary systems, with
recent estimates of binarity for O-type stars of $\gtsimm 70\%$
\citep{Chini:2012, Sana:2012}. In such systems, which consist of two
hot luminous massive stars, the collision of the powerful stellar
winds leads to the formation of high Mach number shocks that emit at
X-ray wavelengths \citep{Stevens:1992}. Historically, colliding winds
binary (CWB) systems have been characterized by high plasma
temperatures and an X-ray over-luminosity (compared to their expected
single star brightness) \citep{Pollock:1987, Chlebowski:1991} with
observational inferences corroborated by theoretical models
\citep{Luo:1990, Stevens:1992, Pittard:1997}. However, more recent
studies examining a wider population and using the {\it XMM-Newton}
and {\it Chandra} satellites indicate that short period WR+O and
O+O-star binary systems have a ratio of $\log(L_{\rm X}/L_{\rm
  bol})\simeq -7$, similar to that expected for single O-stars
\citep{Owocki:1999, DeBecker:2004, Oskinova:2005, Sana:2006,
  Antokhin:2008, Naze:2009, Naze:2011,Gagne:2011,
  Gagne:2012}. Therefore, superlative X-ray brightness - $\log(L_{\rm
  X}/L_{\rm bol})$ as high as -5 - appears to be reserved for the more
massive CWBs with high mass-loss rates (e.g. WR25 -
\citeauthor{Raassen:2003}~\citeyear{Raassen:2003},
\citeauthor{Pollock:2006}~\citeyear{Pollock:2006}; WR140 -
\citeauthor{Pollock:2005}~\citeyear{Pollock:2005}; $\eta$\thinspace
Carinae - \citeauthor{Corcoran:2005}~\citeyear{Corcoran:2005},
\citeauthor{Corcoran:2010}~\citeyear{Corcoran:2010}).

\begin{figure*}
  \begin{center}
    \begin{tabular}{c}
\resizebox{160mm}{!}{\includegraphics[angle=-90]{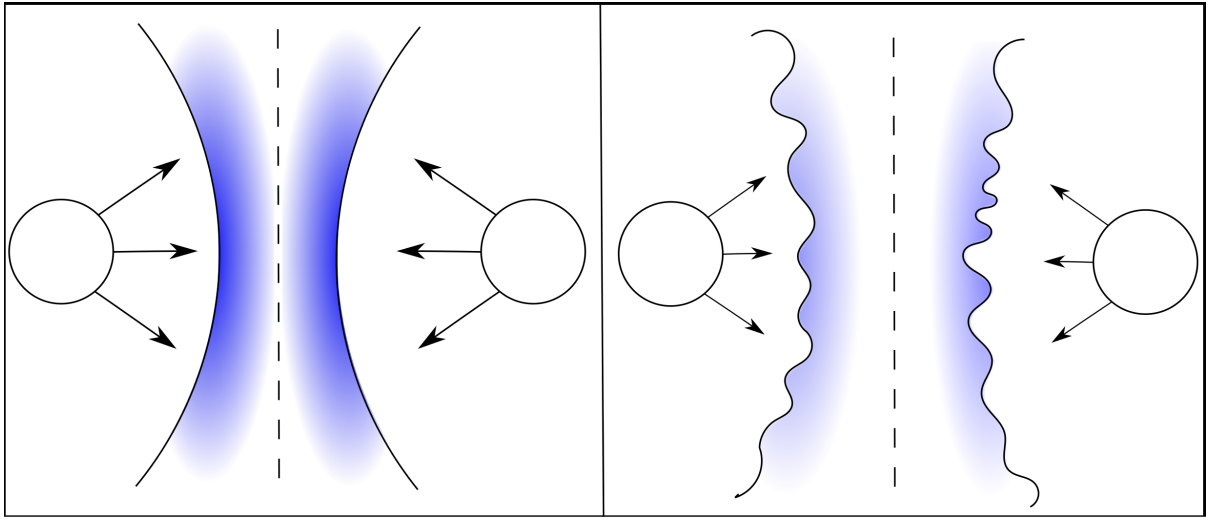}} \\
\end{tabular}
\caption{Cartoon illustration depicting the wind-wind collision in an
  equal winds massive star binary system without (left) and with self
  regulating shocks (right). The stars are represented by the
  circles. Arrows indicate the wind direction, and arrow length
  indicates velocity magnitude. The solid and dashed lines demark the
  regions of post-shock stellar wind and the contact discontinuity,
  respectively. The shaded region indicates plasma temperature -
  fainter shading corresponds to lower temperature.}
    \label{fig:srs_cartoon}
  \end{center}
\end{figure*}

Can current models of CWBs account for the observed spread of over
three orders of magnitude in X-ray luminosity from CWBs
\citep{Gagne:2012}? Specific studies of archetypal systems around the
higher luminosity end of the distribution have yielded promising
results. For example, three dimensional simulations of
$\eta$\thinspace Carinae and WR140, which include orbital motion,
radiative cooling, and in some cases radiative driving, are able to
explain the X-ray lightcurves and spectra reasonably well
\citep{Okazaki:2008, Parkin:2009, Parkin:2011, Russell:2011}. In
contrast, models of WR22 by \cite{Parkin_Gosset:2011} over-predict
\Lx~by up to two orders of magnitude, with the best agreement (a
factor of roughly six over-estimate) achieved when a wind-photosphere
collision occurs and the majority of the X-ray emission is
extinguished. Problems also arise for lower mass CWB systems. A model
of an O6V+O6V binary by \cite{Pittard:2009} and
\cite{Pittard_Parkin:2010} revealed an estimated $\log(L_{\rm
  X}/L_{\rm bol})$ of between -6 and -6.3 (depending on the viewing
angle). This should be compared against observed values for systems
with orbital periods of 2-3 days that have $\log(L_{\rm X}/L_{\rm
  bol})$ between -6.2 and -7.3 \citep[][]{Naze:2009,
  Gagne:2011,Gagne:2012}. \cite{Pittard_Parkin:2010} have presented
evidence that this discrepancy may, in part, be due to the spectral
fitting procedure used to extract parameters from observations, which
they show to under-predict the actual X-ray luminosity (which is known
from the models) by up to a factor of two, particularly for short
period systems where occultation may occur. Alternatively, the
inconsistency with observations may indicate that some additional
physics is required in the models.

Consideration of radiative wind driving in a massive star binary
system led to the discovery of two interesting effects: radiative
inhibition \citep{Stevens:1994} and sudden radiative braking
\citep{Owocki:1995, Gayley:1997}. In the former, the acceleration of
the stellar wind may be reduced by the radiation field of the binary
companion, whereas the latter effect concerns hypersonic flows being
effectively halted in their tracks enabling a wind-wind collision in
systems where one would not occur on the basis of a ram pressure
balance alone. One factor that has not been previously studied in the
CWB paradigm is the influence of the ionizing X-rays from the
wind-wind collision shocks on the wind driving. \cite{Stevens:1990}
examined the dependence of the radiative line force on X-ray
ionization for the case of a high-mass X-ray binary system, finding
that the stellar wind acceleration could be significantly suppressed
by a particularly bright compact object because the excessive X-ray
ionization reduces the radiative line force \citep[see
also][]{Stevens:1991}. This effect has also been explored for
line-driven instability shocks embedded in a massive star's wind
\citep{Krticka_Kubat:2009,Krticka:2009} and for radiatively driven
disk winds of active galactic nuclei \citep{Proga:2000}.

In this paper we make the first attempt to examine the feedback of
ionizing X-rays from the wind-wind collision shocks on wind
acceleration in a massive star binary system. Because of the direct
coupling between the radiation force that drives the stellar winds and
the ionizing X-ray emission that results from the wind-wind collision,
we term this effect {\it self regulated shocks}
(SRSs). Fig.~\ref{fig:srs_cartoon} depicts the basic scenario under
consideration and highlights some key effects due to SRSs. Firstly,
wind velocities are reduced (shorter arrows in the right panel) which
causes a lower post-shock plasma temperature (fainter
shading). Consequently, radiative cooling may become sufficiently
important to introduce instabilities which will perturb the shock
fronts \citep{Stevens:1992, Parkin_Pittard:2010, vanMarle:2011,
  Parkin:2011, Lamberts:2011}. The goal of this work is to provide a
qualitative picture, and initial quantitative estimates, of when/if
the SRS effect might be important. Therefore, we will make
simplifications in order to elucidate the physics.

The structure of this paper is as follows: In \S~\ref{sec:line_force}
we calculate the influence of X-ray ionization on the line force due
to an ensemble of spectral lines. The semi-analytical wind
acceleration model is described in \S~\ref{sec:model}, followed by
results for model binary systems in \S~\ref{sec:results}. An
approximate model for SRSs is presented in \S~\ref{sec:approx}. We
compare results to observations in \S~\ref{sec:obs_compare} and then
discuss some implications of our findings, and possible avenues for
going beyond the illustrative wind acceleration model adopted in this
work, in \S~\ref{sec:discussion}. The main conclusions of this work
are summarised in \S~\ref{sec:conclusions}.

\section{The line force}
\label{sec:line_force}

For the wind models that will be presented in \S~\ref{sec:model}, we
need to compute the radiation force due to spectral lines for
appropriate stellar parameters while accounting for the influence of
X-ray irradiation arising from a wind collision.  We will adopt an
approximate treatment of the radiation force due to spectral lines
following \cite*{Castor:1975} (hereafter \cite{Castor:1975}) and
closely follow the approach by \cite{Stevens:1990} to estimate the
effect of X-ray ionization on the line force -- essentially, our goal
is to repeat their calculations for the stellar parameters appropriate
to our study.  In this section we outline the method and the
implementation used here.  For full details of the methodology and
discussions of its validity, we refer the reader to
\cite{Castor:1975}, \cite{Abbott:1982} and \cite{Stevens:1990}.

The total force due to lines is given by,
\begin{equation}
  f_{\rm rad} = \frac{\sigma_{\rm e}F}{c} M(t) \; ,
\end{equation}
where $\sigma_{\rm e}$ is the electron scattering opacity and $F$ is
the radiative flux. $M(t)$ is known as the line force multiplier,
which depends on the dimensionless optical depth parameter in a
stellar wind, defined by

\begin{equation}
t = \sigma_{\rm e} \rho v_{\rm th} \left(\frac{d v}{d r}\right)^{-1} \; , \label{eqn:t}
\end{equation}
where $\rho$ is the mass density, $v_{\rm th}$ is the thermal velocity
of a hydrogen atom and ${d v}/{d r}$ is the radial
velocity gradient. The Sobolev optical depth of a spectral line
between lower state $l$ and upper state $u$ is given by
$\tau^{S}_{u,l} = \eta_{u,l} t$ where

\begin{equation}
 \eta_{u,l} = \frac{h c}{4 \pi} \frac{n_{l} B_{l,u} - n_{u} B_{u,l}}{
   \sigma_{\rm e} \rho v_{\rm th}} \; .
\end{equation}
Here, $n_{l}$ and $n_{u}$ are the lower and upper level population
number densities and $B_{l,u}$ and $B_{u,l}$ are the usual Einstein
coefficients for absorption and stimulated emission, respectively. The
force multiplier, $M(t)$, is composed from a sum over all line
transitions

\begin{equation}
M(t) = \sum_{\rm lines} \Delta\nu_{D} \frac{F_{\nu}}{F} \frac{1 - \exp(-\eta_{u,l} t)}{t} \; ,
\end{equation}
where $\Delta\nu_{D}$ is the Doppler width and $F_{\nu}$ is the
specific flux at the line frequency ($\nu$).

To evaluate $M(t)$, we need to supply a list of line transitions
(frequencies and oscillator strengths), specify the form of the
radiation field $F_{\nu}$, and compute the associated level
populations ($n_{l}$, $n_{u}$, relative to the total density $\rho$).

The line list used in this study is drawn from two sources. For
low-ionization metal atoms/ions, we use the CD23 line database of
\cite{Kurucz:1995}. From this source we include elements with atomic
number $6 \leq Z \leq 30$ and include ionization stages {\sc i} --
{\sc v} with the following exceptions: for C, we include only {\sc i}
-- {\sc iv} while for $Z > 20$ we include ions {\sc i} -- {\sc vii},
where available. In order to extend our calculations to regimes of
higher ionization, we also included data from the {\sc chianti} atomic
database \citep{Dere:1997,Dere:2009}. From this source, we take line
lists for H and He and the high ions of the astrophysically abundance
metals: C, N, O, Ne, Mg, Si, S, Ar, Ca, Fe and Ni (for each of these
metal, we include {\sc chianti} line lists for all available ions that
we did not take from Kurucz \& Bell 1995; we excluded theoretically
predicted lines from the database). In total, our line list contains
$\sim 7.7 \times 10^5$ transitions.

The stellar radiation field, $F_{\nu}$ was taken from ATLAS9 model
atmosphere grids \citep{Castelii:2004}. For the specific stellar
parameters used, see below.

The level populations ($n_{l}$, $n_{u}$) for each transition were
computed in a two stage process. First we used Cloudy v10.00
\citep[][]{Ferland:1998} to compute the ionization stage of a shell of
gas illuminated by a specified radiation field. In all cases, we
assumed that the irradiating spectrum contains two components:
emission from the star and hard radiation associated with emission
from the wind collision region. The shape of the stellar component was
taken from the same model atmospheres used for $F_{\nu}/F$. In setting
the intensity of this component, we follow \cite{Stevens:1990} and
consider only a single value for the ratio of the electron number
density to the geometrical dilution factor ($n_{e} / W = 3.5 \times
10^{10}$~cm$^{-3}$). To describe the spectral shape of the hard
ionizing radiation, we adopt a thermal Bremsstrahlung spectrum at a
temperature of 10~keV. The intensity of this component is specified as
an ionization parameter,
\begin{equation}
  \xi = \frac{4 \pi F_{\rm X}\mu m_{\rm H}}{\rho} \label{eqn:xi}
\end{equation}
where, in this work, $F_{\rm X}$ is the flux of X-rays from the wind
collision shocks, and $\rho$ is the gas density. The value of $\xi$ is
varied to quantify the affect of X-ray ionization on $M(t)$.
From the ion populations provided by the Cloudy calculations, we
compute level populations assuming local thermodynamic equilibrium
(LTE, adopting the gas temperature calculated by Cloudy). Although
simplistic, this assumption makes it easy to compute the force
multiplier reasonably quickly. Ideally, full non-LTE calculations
should be performed for complete atomic models associated with each
ion. This, however, would significantly complicate the calculation and
is not expected to qualitatively affect our findings \citep[see][for
further discussion]{Stevens:1990}.

\subsection{Example calculation}
\label{sect:example_M}

\begin{figure}
  \begin{center}
    \begin{tabular}{c}
\resizebox{85mm}{!}{\includegraphics{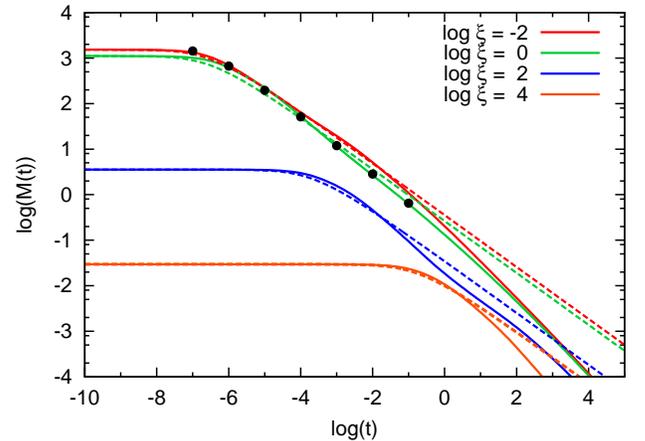}} \\
  \end{tabular}
  \caption{Force multiplier versus dimensionless optical depth
    parameter ($t$) for $\log \xi = $ -2, 0, 2 and 4 (solid
    lines). Dashed lines show our fits (see text). These calculations
    are for a star with $T_{\rm eff} = 38500$~K, $\log g = 3.92$ and
    $\log (Z / Z_{\odot}) = 0$. For comparison, we also show
    calculations from Abbott (1982) for a star with $T_{\rm eff} =
    40000$~K and $\log g = 4.0$ (black spots).}
    \label{fig:Mplot}
  \end{center}
\end{figure}

Using the procedure outlined above, we can calculate $M(t)$ accounting
for the effects of excess ionization (as controlled by $\xi$). As an
example, we show results for a star with effective temperature $T_{\rm
  eff} = 38500$~K, surface gravity $\log g = 3.92$ and solar
metallicity $\log (Z / Z_{\odot}) = 0$ in Fig.~\ref{fig:Mplot}.

As expected, our results are in good agreement with
\cite{Stevens:1990}. For each value of $\xi$, $M$ is largest (and
constant) when $t$ is sufficiently small that all lines are optically
thin. At large $t$, $M$ decreases as lines become optically thick and
saturate. For calculations with low ionization parameter ($\log \xi <
0$), $M$ remains significant ($M \gtsimm 1$) up to around $t \sim 1$
($M$ is essentially independent of ionization parameter for $\log \xi
< -2$). As found by \cite{Stevens:1990}, we also see that as $\log
\xi$ is increased beyond zero, $M$ drops and the regime in which
$M(t)$ is well-described by the optically thin limit extends to higher
$t$-values. For $\log \xi > 3$, the force multiplier is always small.

For comparison, we also show in Fig.~\ref{fig:Mplot} the $M$-values
reported by \cite{Abbott:1982} from calculations for a star with
similar parameters ($T_{\rm eff} = 40000$~K, $\log g = 4.00$ and
$n_{e}/W = 1.8 \times 10^{11}$~cm$^{-3}$). Since no excess ionization
radiation was included by \cite{Abbott:1982}, his calculations should
be compared to our results for the lowest ionization parameter shown
($\log \xi = -2$). In general, the agreement is very good -- the
biggest discrepancy occurs around $\log t = -2.5$ and is at worst a
factor of two.

\subsection{Parameterizing the force multiplier}

Although the force multiplier $M$ can be directly used to specify the
line force, it is convenient to parametrize its dependence on $t$ for
use in wind calculations. Although this approach means that the full
complexity of $M(t)$ is not captured, it is widely used because of the
relative ease of manipulating simply-parametrized forms for $M(t)$
when deriving wind solutions.

The basic ansatz under the \cite{Castor:1975} approximation is to fit
a power-law to the run of $M(t)$ with $t$,
\begin{equation}
  M(t) = k t^{-\alpha} \label{eqn:basicMt}
\end{equation}
where $\alpha$ defines the slope and $k$ the amplitude of $M$ at $t=1$
(i.e., $k=M(1)$). To capture the flattening of $M(t)$ for small $t$,
we follow \cite{Owocki:1988}, and modify Eq~(\ref{eqn:basicMt}) such
that the force multiplier becomes constant at low $t$ (as it must in
the optically thin limit),
\begin{equation}
  M(t,\xi) = k(\xi) t^{- \alpha} \left[\frac{(1 +
      \tau_{\rm max})^{1-\alpha}-1}{\tau_{\rm max}^{1 - \alpha}}\right] \label{eqn:Mtxi}
\end{equation}
where $\tau_{\rm max} = \eta_{\rm max}(\xi) t$. 
In Eq~(\ref{eqn:Mtxi}) we now explicitly indicate that $M$ depends on
both $t$ and $\xi$. Throughout this paper, we will choose to describe
the influence of $\xi$ on $M$ via the CAK parameters $k(\xi)$ and
$\eta_{\rm max}(\xi)$. Allowing for $\xi$-dependence in these
quantities captures the two systematic changes in $M(t, \xi)$ with
$\xi$: decreasing $\eta_{\rm max}(\xi)$ with increasing $\xi$ allows
the turnover in $M(t)$ to shift to higher $t$ with increasing $\xi$,
while a reduction in $k(\xi)$ at large $\xi$ describes the overall
decrease in $M$ for larger ionization parameters.

We follow \cite{Stevens:1990} in {\it choosing} that $\alpha$ does not
vary with $\xi$. Although it is certainly possible to allow $\alpha$
to vary, this mostly just adds unwarranted complexity to the
parametrization. As is clear from Fig.~\ref{fig:Mplot}, the slope of
$\log M$ versus $\log t$ is not constant, meaning that a best-fit
$\alpha$ is in any case a function of the range across which it is
fit. Therefore, in all of our calculations, we do not fit $\alpha$ but
rather fix it to the value that is required in order to reproduce the
correct observed terminal velocity for a single star of the
appropriate spectral type in a standard CAK theory\footnote{The
  terminal velocity computed in a single star wind calculation does
  also depend on $k$, but to a much lesser extent than $\alpha$, as
  one would expect from the functional form of $M(t)$.}. For the
example star discussed in Section~\ref{sect:example_M}, this is
$\alpha = 0.57$.

With $\alpha$ fixed, we derive values of $k(\xi)$ and $\eta_{\rm
  max}(\xi)$ by fitting our computed $M(t)$ curves to the functional
form given by Eq~(\ref{eqn:Mtxi}). We restrict this fitting to $\log t
< 0$, the regime in which $M(t)$ is expected to be dynamically
significant. To illustrate the accuracy of this approach, the derived
fits from our example calculation are over-plotted in
Fig.~\ref{fig:Mplot}. As expected, the fits are always very good in
the optically thin limit and generally agree to within a few tens of
per cent across the range of interest (i.e. when $M \gtsimm
1$). However, there are clear imperfections, particularly in cases
where the slope of $M(t)$ deviates from a constant power law (e.g. in
our $\log \xi = 2$ case). Nevertheless, the parametrized form provides
a convenient description and reproduces the force multiplier to within
a factor of two, which is adequate precision for the purpose of this
investigation. We provide tabulated values of $k(\xi)$ and $\eta_{\rm
  max}(\xi)$ in the Appendix.
\subsection{Rescaling of the force multiplier}
\label{subsec:rescale}

As mentioned in the previous section, in fitting $k(\xi)$ and
$\eta_{\rm max}(\xi)$ we specified the value of $\alpha$ {\it a
  priori} with the aim that the resulting $M(t,\xi)$ produced a
terminal wind velocity in agreement with observed values. We now also
rescale the force multiplier $M(t,\xi)$ to produce a wind {\it
  mass-loss rate} in agreement with observed values, which equates to
multiplying $k(\xi)$ by a correction factor\footnote{The correction
  factors are 0.74 and 0.34 for the O6V and O4III stars,
  respectively.}. At the cost of some subjective rescaling, this
approach has the advantage of ensuring that the line force used in the
colliding-winds model in the following section will produce sensible
wind parameters while also allowing the influence of X-ray irradiation
to be explored. We note that this modification is of smaller magnitude
than the current uncertainties in mass-loss rates and wind
acceleration in massive stars \citep[see][for a recent
review]{Puls:2008}.

\subsection{Results of the line force calculation}
\label{subsec:line_force_results}

\begin{figure}
  \begin{center}
    \begin{tabular}{c}
\resizebox{80mm}{!}{\includegraphics{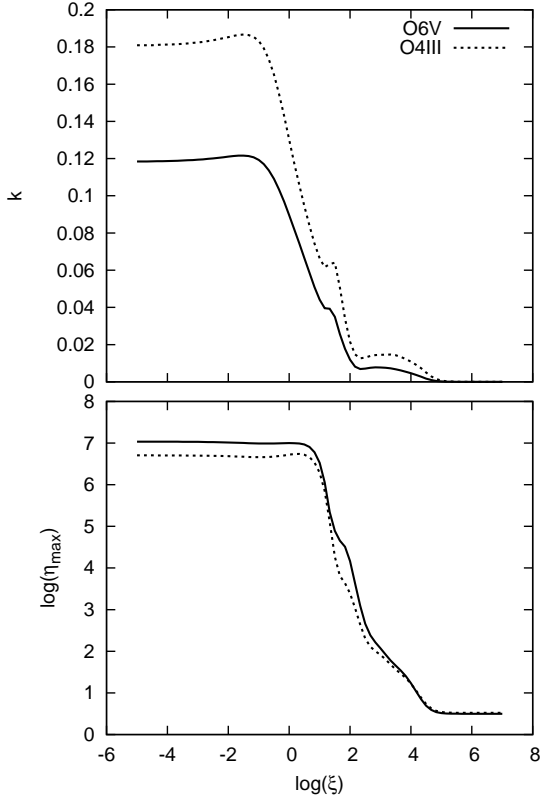}} \\
 \end{tabular}
 \caption{Plots of the line force parameters $k$ (upper) and
   $\log(\eta_{\rm max})$ (lower) as a function of ionization
   parameter, $\xi$.}
    \label{fig:k_etamax}
  \end{center}
\end{figure}

Line force calculations were performed for two massive stars: O6V and
O4III (see Table~\ref{tab:stellar_models} for full sets of stellar
parameters). In Fig.~\ref{fig:k_etamax} we show the resulting $k(\xi)$
and $\eta_{\rm max}(\xi)$.  Clearly, for $\log(\xi)>0$ the line force
is effectively suppressed, whereas for $\log(\xi)<0$ ionization
effects are negligible and radiative acceleration is largely
unaffected. 

\begin{table*}
\begin{center}
  \caption[]{Parameters used for stellar model atmospheres} \label{tab:stellar_models}
\begin{tabular}{lllllllllll}
\hline
Model & $T_{\rm eff}$ & $M_{\ast}$ & $R_{\ast}$ & $\log(L/L_{\odot})$
& $\log (g)$ & $Z$ & $\dot{M}$ & $v_{\infty}$ & $k(\xi=0)$ & $\alpha$\\
 & $(K)$ & $(M_{\odot})$ & $(R_{\odot})$ & & & & (\Msolpyr) & (\kmps)
 & & \\
\hline
O6V & 38500 & 31.7 & 10.2 & 5.3 &  3.92 & 1 & $2\times10^{-7}$&
2530 & 0.12 & 0.57 \\
O4III & 41500 & 48.8 & 15.8 & 5.8 &  3.73 & 1 & $5\times10^{-6}$&
2750 & 0.18 & 0.63\\
\hline
\end{tabular}
\end{center}
\end{table*}
\section{The wind model}
\label{sec:model}

\subsection{Radiatively driven winds}
\label{subsec:winds}

To compute the wind acceleration we follow
\cite{Stevens:1992}. Alterations have been made to couple the model
with a means of estimating the X-ray luminosity from the wind-wind
collision shocks, and then allow for an ionization parameter ($\xi$)
dependence of the line force. The calculation proceeds by solving for
the wind of one of the stars, with the influence of the companion star
appearing in the effective gravitational potential and as a
contribution to the total radiative flux in the line
force. Subsequently, we change to the frame of reference of the
companion star and solve for its wind in an equivalent manner. In the
following we describe the solution procedure for each wind. In
\S~\ref{subsec:shocks} we discuss how to infer the shock properties
from the two wind solutions.

To simplify the problem we consider steady state solutions for the
flow along the line-of-centres between the stars with the forces
arising due to orbital motion ignored\footnote{\cite{Parkin:2011}
  considered the effect of centrifugal acceleration due to orbital
  motion on the wind acceleration and found that it made a correction
  of a few per cent. Furthermore, for the systems considered in this
  paper, the wind speeds are sufficiently large compared to the
  orbital velocities that there should be no significant offsets in
  the position of the wind-wind collision region due to orbital motion
  \citep{Parkin:2008}.}.  We assume that the flow is symmetric about
the line of centres and that the wind flows purely radially from the
star. We also assume that the wind is isothermal with temperature,
$T=0.8~T_{\rm eff}$, where $T_{\rm eff}$ is the effective stellar
surface temperature. Lastly, we do not consider stellar radiation
reflected from the opposing star's photosphere. Some approximate
expressions are used in the model to keep the calculations
straightforward, whilst achieving an accuracy at the order unity
level, and in \S~\ref{sec:discussion} we discuss possible
alternatives. The equations for mass and momentum conservation in the
wind are,
\begin{eqnarray}
\oint \rho {\bf v}\cdot d{\rm S} &=& 0 \rightarrow \dot{M}_{\Omega} = r^2 \rho v\\
  F(r, v, dv/dr) &=& \left(1 - \frac{a^2}{v^2}  \right)v\frac{dv}{dr} +
  \frac{d \Phi}{dr} - \frac{2 a^2}{r} - g_{\rm rad}, \\
  F(r, v, dv/dr) &=& 0, \nonumber
\end{eqnarray}
\noindent where $r$ is the distance along the line of centres measured
from the centre of star 1, $v$ is the wind velocity along the line of
centres, $\rho$ is the density, $\dot{M}_{\Omega}$ is the mass-loss
rate per steradian and $a$ is the isothermal speed of sound. The
gravitational potential due to both stars,
\begin{equation}
 \Phi = -\frac{G M_{\ast1}(1 - \Gamma_{1})}{r} -  \frac{G M_{\ast2}(1 -
   \Gamma_{2})}{d_{\rm sep} - r},
\end{equation}
where $M_{\ast1}$ and $M_{\ast2}$ are the respective masses of star 1
and star 2, $\Gamma_{1}$ and $\Gamma_{2}$ are the respective Eddington
ratio for each star ($\Gamma_{\rm i} = \sigma_{\rm e}L_{\ast \rm i}/4
\pi G M_{\ast \rm i}c$), $d_{\rm sep}$ is the separation of the stars
(measured between their centres), and $G$ is the gravitational
constant. The combined radiative line force from both stars, $g_{\rm
  rad}$ takes the form,
\begin{equation}
  g_{\rm rad} = \frac{\sigma_{\rm e}M(t,\xi)}{c}(F_1 K_1 - F_2 K_2), \label{eqn:RI}
\end{equation}
\noindent where $F_1$, $F_2$ are the radiative fluxes, and $K_1$,
$K_2$ are the finite disk correction factors (FDCFs) for stars 1 and
2, respectively. $M(t,\xi)$ is the line force multiplier
(Eq~\ref{eqn:Mtxi}), which in our formulation has a dependence on both
optical depth, $t$ (Eq~\ref{eqn:t}) and the ionization parameter,
$\xi$ (Eq~\ref{eqn:xi}).

The FDCF is a multiplicative factor used to correct the point source
approximation for the finite size of the stellar disk
\citep{Castor:1974, Castor:1975, Pauldrach:1986}. For our adopted
geometry and assumptions about the flow along the line of centres, we
have,
\begin{eqnarray}
  K_{\rm i}(r,v,dv/dr) & = & \frac{(1 + \sigma_{\rm i})^{1 + \alpha} - (1 +
    \sigma_{\rm i}\mu_{\ast i}^2)^{1 + \alpha}}{\sigma_{\rm i}(1 + \alpha)(1 +
    \sigma_{\rm i})^{\alpha}(1 - \mu_{\ast i}^{2})}, \label{eqn:dynamical_fdcf}
\end{eqnarray} 
where $\mu_{\rm i} = \cos\theta_{\rm i}$ with $\theta_{\rm i}$ being
the angle subtended by the respective stellar disk {\it viewed from a
  point in the wind}, $\mu_{\ast 1}^2 = 1 - R_{\ast 1}^2/r^2$ and
$\mu_{\ast 2}^2 = 1 - R_{\ast 2}^2/(d_{\rm sep} - r)^2$, and $\sigma_{1} =
(r/v)(d v/d r) - 1$ and $\sigma_{2} = (d_{\rm sep}-r )/v)(d v/d r) -
1$. Following \cite{Stevens:1994} and \cite{Pauldrach:1986} we
approximate the FDCFs as purely radial functions and neglect any
velocity or velocity gradient terms. In this limit,
\begin{eqnarray}
K_{\rm i}(r,v,dv/dr) \rightarrow K_{\rm i}(r) = \frac{1 - [1 - \mu_{\ast
    \rm i}^2]^{1 + \alpha}}{(1 + \alpha)(1 -  \mu_{\ast
    \rm i}^2)}.
\end{eqnarray}

In the model considered here, we enforce monotonicity in the flow by
ensuring $dv/dr = \max(dv/dr,0)$, which ensures that the optical depth
parameter, $t$ is a positive valued scalar variable. It follows that
our models do not permit radiative braking (which requires an
inflection in the velocity gradient). \cite{Gayley:1997} comment that
correctly accounting for non-monotonicity in the FDCF allows radiative
braking. Another way of viewing this is that radiative braking
involves a bridging between two monotonic flows, which is not
facilitated by standard \cite{Castor:1975} theory.

To proceed, we make a coordinate transform using the substitution of
variables \citep{Abbott:1980},
\begin{eqnarray}
  u &=& \frac{-2 GM_1(1 - \Gamma_1)}{r a^2};  \\
  w &=& \frac{v^2}{a^2}; \\
  w' &=& r^2 v \frac{dv}{dr} [G M_1 (1 -\Gamma_1)]^{-1},
\end{eqnarray}
\noindent leading to
\begin{equation}
  F(u, w, w') = \left(1 - \frac{1}{w} \right)w' + h(u) - g E(w',\xi)
  B(u)w'^{\alpha}, \label{eqn:Fuww'}
\end{equation}
\noindent where
\begin{equation}
  g = \Gamma_{1}/(1 - \Gamma_{1}),
\end{equation}
\begin{equation}
 E(w',\xi) = k(\xi) C^{-\alpha} w' \left[\left(\frac{1}{\eta_{\rm max}
     C} +\frac{1}{w'}\right)^{1-\alpha} - (\eta_{\rm max} C)^{\alpha -1}\right],
\end{equation}
\begin{equation}
  C = \frac{\sigma_{\rm e} \dot{M}_{\Omega} v_{\rm th}}{G M_{\ast 1}(1
    - \Gamma_{1})}, \label{eqn:C}
\end{equation}
\begin{equation}
  B(u) = K_1(u) - \left(\frac{M_{\ast 2} \Gamma_2}{M_{\ast 1} \Gamma_1} \right)K_{2}(u)A(u),
\end{equation}
\begin{equation}
  h(u) = 1 + \frac{4}{u} - \frac{M_{\ast 2}(1 - \Gamma_2)}{M_{\ast 1}(1 - \Gamma_1)}A(u),
\end{equation}
\noindent and
\begin{equation}
 A(u) = \left(\frac{u_{\rm d}}{u - u_{\rm d}} \right)^2.
\end{equation}
\noindent (Note that our definition of $C$ in Eq~(\ref{eqn:C}) differs
from \cite{Stevens:1994}'s equation (13)). The FDCFs
\begin{equation}
  K_{1}(u) = \frac{1 - \left[1 - (u/u_{\ast
            1})^2\right]^{1 + \alpha}}{(1 +
    \alpha) (u/u_{\ast 1})^{2}}, \label{eqn:FDCF_RI1}
\end{equation}
\noindent and 
\begin{equation} 
  K_{2}(u) = \frac{1 - \left[1 - (u/u_{\ast
        2})^2A(u)\right]^{1 + \alpha}}{(1 +
    \alpha) (u/u_{\ast 2})^2 A(u)}, \label{eqn:FDCF_RI2}
\end{equation}
\noindent where $u_{\rm d} = u(d_{\rm sep})$, $u_{\rm \ast \rm i} =
u(R_{\ast \rm i})$.
The strategy for finding a consistent wind solution is centered around
the use of the critical point conditions,
\begin{eqnarray}
  f_{1}(w_{\rm c},w'_{\rm c},C_{\rm c}) &=& F(u,w,w') = 0, \label{eqn:f1}\\
  f_{2}(w_{\rm c},w'_{\rm c},C_{\rm c}) &=& \frac{\partial F}{\partial w'} = 0, \label{eqn:f2}\\
  f_{3}(w_{\rm c},w'_{\rm c},C_{\rm c}) &=& \frac{\partial F}{ \partial
    u} + w'\frac{\partial F}{\partial w} = 0. \label{eqn:f3}
\end{eqnarray}
Eqs~(\ref{eqn:f1})-(\ref{eqn:f3}) are, respectively, the equation of
motion, the singularity condition, and the regularity condition. The
subscript ``c'' denotes the value of the given parameter at the
critical point. Our set of equations differs slightly from those of
\cite{Stevens:1994}. Specifically, we lack the singular presence of
the eigenvalue of the problem (namely the mass-loss rate). Therefore,
we cannot use the equations for $f_{1}$, $f_{2}$, and $f_{3}$ to
derive closed form relations for $w_{\rm c}$ and $w'_{\rm
  c}$. Instead, noting that Eqs~(\ref{eqn:f1})-(\ref{eqn:f3}) compose
three equations in three unknowns, we solve for $w_{\rm c}$, $w'_{\rm
  c}$, and $C_{\rm c}$ using a multi-dimensional root finder
\citep[see, e.g.,][]{Press:1986}, where the critical point conditions
derived by \cite{Stevens:1994} are used as the initial guess.

\subsection{The post-shock winds}
\label{subsec:shocks}

Once the wind profiles have been calculated we proceed to estimate the
X-ray luminosity from the individual wind collision shocks. The
separate values are then combined to evaluate the total shocked-wind
X-ray luminosity. The final step is to use the estimate of the
intrinsic X-ray luminosity from the shocked winds (Eq~\ref{eqn:lx}) to
evaluate the ionization parameter, $\xi$ (Eq~\ref{eqn:xi}). (Note that
when we swap from the frame of reference of one of the stars to its
companion's, we interchange the indices in Eqs~\ref{eqn:lx} and
\ref{eqn:zeta}).

We approximate each shock as a thin shell, and take the pre-shock wind
velocity and density to be $v_{\rm sh}=v(r_{\rm bal})$ and $\rho_{\rm
  sh}=\rho(r_{\rm bal})$, respectively, where $r_{\rm bal}$ is the ram
pressure balance point. The {\it mean} post-shock gas temperature
(i.e. averaged over the bow shock), $T_{\rm ps}$, can be estimated
from the Rankine-Hugoniot shock jump conditions, $kT\simeq \frac{1}{2}
\times1.17 (v_{\rm sh}/10^{8}$cm~s$^{-1})^{2}$~keV, where the factor
of a half is a correction to account for shock obliquity. The
0.01-10~keV X-ray luminosity from the respective wind-wind collision
shocks is then estimated using the simple relation:
\begin{eqnarray}
  L_{\rm X i} = \frac{1}{2}\dot{M_{\rm i}}v_{\rm i}^{2} \Xi_{\rm i}\left(\frac{1}{1+\chi}\right), \label{eqn:lx}
\end{eqnarray}
\noindent where the total mass-loss rate is approximated as
$\dot{M}\approx 4 \pi \dot{M}_{\Omega}$. (Note that the X-ray emitting
region of the shocks is taken to be a point source situated at the ram
pressure balance point along the line-of-centres). The parameter $\Xi$
approximates the thermalization of wind kinetic power. In
\S~\ref{sec:results} we consider models of wind-wind collision and
wind-photosphere collision (as a result of the stronger wind
overwhelming the weaker wind). In the latter circumstance we take
$\Xi$ to be the fractional solid angle subtended by the disk of the
companion star,
\begin{equation}
  \Xi = \frac{1}{2}(1 -\epsilon_{\ast}), \label{eqn:xi_wind_phot}
\end{equation}
where $\epsilon_{\ast}^2=1- (R_{\ast 2}/d_{\rm sep})^2$. For a
wind-wind collision we take $\Xi$ to be the fractional wind kinetic
power normal to the contact discontinuity \citep{Zabalza:2011},
\begin{equation}
  \Xi = \frac{1}{4}\left(\frac{\pi \zeta_{\rm eff}}{1+\zeta_{\rm eff}}
  \right)^2, \label{eqn:zabalza_xi}
\end{equation}
where the effective wind momentum ratio of the system,
\begin{equation}
 \zeta_{\rm eff}=\frac{\dot{M}_{2} v_{\rm sh 2}}{\dot{M}_{1} v_{\rm sh
     1}}. \label{eqn:zeta}
\end{equation}
The cooling parameter, $\chi$ appearing in Eq~(\ref{eqn:lx}) derives
from the ratio of the characteristic flow time to the cooling time
\citep{Stevens:1992},
\begin{equation}
\chi = \left(\frac{v_{\rm sh}}{10^8 \cmps}\right)^4
\left(\frac{d_{\rm sep}}{10^{12}\;{\rm cm}}\right)
\left(\frac{\dot{M}}{10^{-7}\Msolpyr}\right)^{-1}. \label{eqn:chi}
\end{equation}
If $\chi \ltsimm 1$ the post-shock gas is radiative, whereas if $\chi
\gg 1$ the post-shock gas is adiabatic. It is useful to note the
different scalings of \Lx\ as the importance of cooling changes. For
adiabatic shocks we have $\Lx \propto (\dot{M}/v_{\rm sh})^2$,
therefore a decrease in $v_{\rm sh}$ leads to an increase in \Lx. In
contrast, when the shocks are radiative $\Lx \propto \dot{M}v_{\rm
  sh}^2$, and a decrease in $v_{\rm sh}$ reduces \Lx.

In \S~\ref{sec:results} we consider calculations with either an
attenuated or unattenuated X-ray flux. For the latter we merely have,
$F_{\rm X}(r)=(L_{\rm X 1} + L_{\rm X 2})/4 \pi (r_{\rm
  bal}-r)^2$. For the former, an attenuated flux is calculated by
first scaling a 0.01-10 keV X-ray spectrum - derived from the MEKAL
plasma code \citep{Kaastra:1992, Mewe:1995} - such that its total
luminosity matches the value from Eq~(\ref{eqn:lx}). The spectra from
both winds are then combined, and the column density of gas upstream
of the shock, $N_{\rm H} = \int^{r_{\rm bal}}_r \rho(r) dr$, is used
to attenuate the spectrum. (Absorption due to the post-shock layers is
neglected.)  Finally, the resulting spectrum is integrated to acquire
the attenuated luminosity, $L_{\rm X att}$ from which the ionization
parameter can be determined. To this end we use version $c08.00$ of
Cloudy \citep[][see also
\citeauthor{Ferland:1998}~\citeyear{Ferland:1998}]{Ferland:2000} to
calculate the opacity.

To test the accuracy of our model, we made a calculation for an
O6V+O6V binary at a separation of 30\Rsol and compared the estimated
X-ray luminosity to model cwb1 from \cite{Pittard:2009} and
\cite{Pittard_Parkin:2010} (a 3D hydrodynamical model with radiatively
driven winds). We found that our model over-predicted the intrinsic
0.1-10 keV X-ray luminosity by a factor of roughly two. Therefore, for
all models examined in this paper we multiply Eq~(\ref{eqn:lx}) by a
factor of 1/2. Furthermore, when computing $\log(L_{\rm X}/L_{\rm
  bol})$ from our models we define $L_{\rm X}$ as the 0.5-10 keV X-ray
luminosity to be consistent with observational studies \citep[see, for
example,][]{Naze:2009, Naze:2011, Gagne:2011}.

\begin{figure}
  \begin{center}
    \begin{tabular}{c}
\resizebox{80mm}{!}{\includegraphics{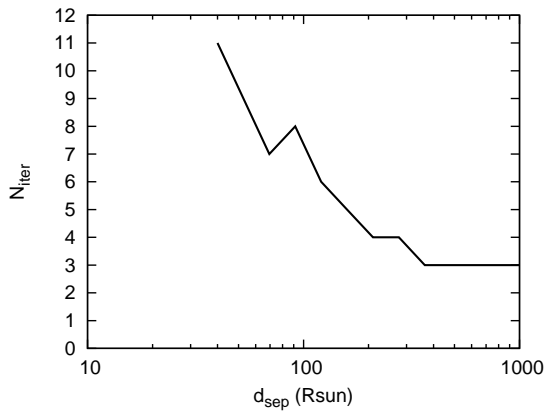}} \\
 \end{tabular}
 \caption{The number of iterations, $N_{\rm iter}$ required to reach
   convergence in the O6V+O6V binary model calculations.}
    \label{fig:vel_iter}
  \end{center}
\end{figure}

\subsection{Solution strategy}
To summarise, the steps in the calculation are:
\begin{enumerate}
\item Begin by setting $\xi=0$ everywhere.
\item Compute $v(r)$ and $\rho(r)$ using the wind acceleration model
  described in \S~\ref{subsec:winds} for the respective
  stars. \label{step:wind}
\item Determine the ram pressure balance point between the winds and
  use this to find $L_{\rm X}$ for both post-shock winds. For a
  wind-photosphere collision, the balance point is taken to be at the
  surface of the companion star. \label{step:ram}
\item Calculate the X-ray ionization parameter, $\xi(r)$ (see
  \S~\ref{subsec:shocks}). \label{step:xi}
\item Calculate the change in $v_{\rm sh}$ relative to the last
  iteration and if convergence is not achieved\footnote{For the
    calculations presented in this paper we required the fractional
    difference in $v_{\rm sh}$ (summed over both winds) between
    consecutive iterations to be $\leq 10^{-4}$.} then repeat steps
  (\ref{step:wind})-(\ref{step:xi}).
\end{enumerate}
Fig.~\ref{fig:vel_iter} shows the number of iterations required to
reach convergence. A larger number of iterations are required for
smaller separations where the affect of SRSs is greatest.

\section{Results}
\label{sec:results}

\begin{figure}
  \begin{center}
    \begin{tabular}{c}
\resizebox{80mm}{!}{\includegraphics{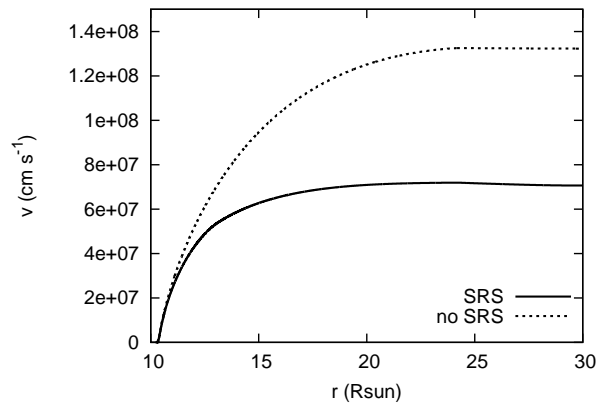}} \\
 \end{tabular}
 \caption{Wind velocity as a function of radius for the O6V binary
   with a separation of $d_{\rm sep}=60~R_{\odot}$. The different
   curves show the consecutive iterations. Convergence is reached
   after $\sim 5$ iterations.}
    \label{fig:standard_vel}
  \end{center}
\end{figure}

In this section we examine the influence of the ionizing X-rays from
the wind-wind collision shocks on the resulting wind acceleration. We
have constructed two binary systems which we use to explore the impact
of self-regulating shocks across a small range of spectral types: an
O6V+O6V binary and an 04III+O6V binary. We also consider the collision
of the O4III star's wind against the photosphere of the O6V star,
which arises at binary separations smaller than 300\Rsol~in the
O4III+O6V case. In the following sections we first examine the general
properties of the self-regulating shocks scenario using the O6V+O6V
binary system as our fiducial test case, then consider the importance
of SRSs as a function of stellar separation for the different model
binaries.

\subsection{General properties}
\label{subsec:general}

\begin{figure}
  \begin{center}
    \begin{tabular}{c}
\resizebox{80mm}{!}{\includegraphics{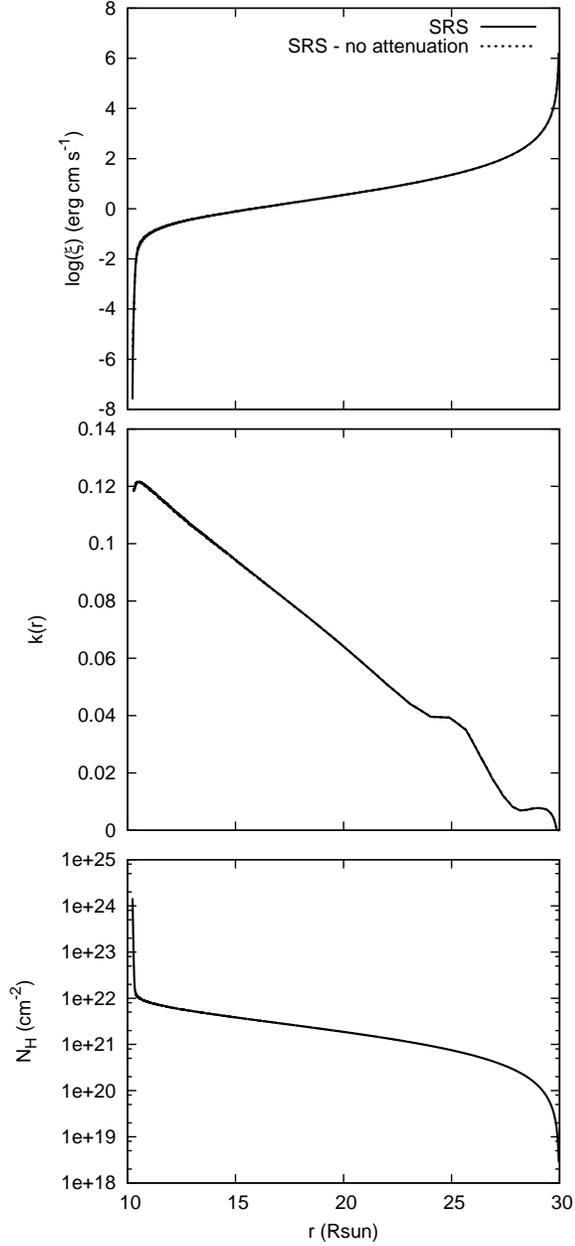}} \\
 \end{tabular}
 \caption{Radial profiles of $\xi$ (upper), $k(r)$ (middle), and the
   total column density measured from the shock through the wind,
   $N_{\rm H}$ (lower). }
    \label{fig:general_plots}
  \end{center}
\end{figure}

\begin{figure}
  \begin{center}
    \begin{tabular}{c}
\resizebox{80mm}{!}{\includegraphics{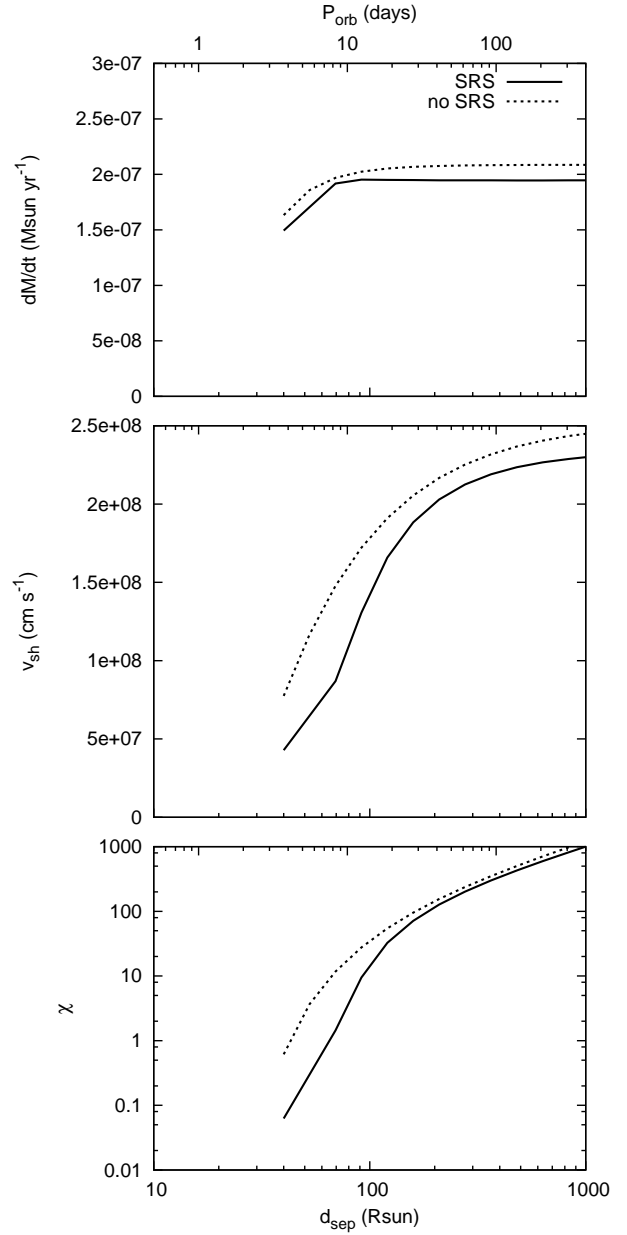}} \\
 \end{tabular}
 \caption{Comparison of calculations for the O6V+O6V binary with and
   without self regulating shocks. Orbital periods are calculated
   assuming circular orbits. From top to bottom: $\dot{M}$, $v_{\rm
     sh}$, and $\chi$.}
    \label{fig:general_plots2}
  \end{center}
\end{figure}

We begin by examining the O6V+O6V binary at a separation of $d_{\rm
  sep}=60\Rsol$. Fig.~\ref{fig:standard_vel} shows the dramatic
influence of SRSs on the wind acceleration compared to a calculation
without this effect included. Three results are immediately apparent
from this plot: i) the pre-shock velocity is considerably reduced, ii)
the wind acceleration is inhibited, and, iii) the acceleration region
is smaller.

What causes such a significant difference between wind calculations
with and without SRSs? Fig.~\ref{fig:general_plots} shows the
variation of $\xi$ with radius from the star. Close to the shocks
(which reside at $r=30\Rsol$ in this example), $\log(\xi) \gg 0$ which
is sufficient to strongly suppress the line force
(\S~\ref{sec:line_force}). In fact, throughout most of the wind the
value of $\xi$ is large enough that the wind acceleration, $g_{\rm
  rad} \propto k(\xi)$ will be inhibited somewhat (see
Fig.~\ref{fig:k_etamax}), and this becomes clear when one examines the
run of $k(\xi)$ against radius. We note, however, that the sharp rise
in $\xi$ close to the shocks is an unphysical consequence of
concentrating all of the X-ray emission from post-shock winds at the
stagnation point. SRSs are important when $\xi$ reaches relatively
high values in the wind acceleration region (i.e. well away from the
shocks) and, therefore, the spike seen in the top panel of
Fig.~\ref{fig:general_plots} does not affect the main conclusions of
this work.

For comparison, we also show in Fig.~\ref{fig:general_plots} (top
panel) a calculation performed with an unattenuated X-ray flux, which
differs only very slightly from the calculation with an attenuated
X-ray flux. Although the total accrued column density, $N_{\rm H tot}$
steadily increases when tracking back from the shocks towards the star
(lower panel of Fig.~\ref{fig:general_plots}) it never reaches a
sufficiently high value to impact the X-ray flux from the wind-wind
collision shocks. Therefore, the decrease in $\xi$ as one moves away
from the shocks results from an increase in the wind density and
geometrical dilution of the X-rays. This result holds true for all of
the models considered in this work. However, the influence of
attenuation may become more important for higher mass-loss rates
and/or when the winds contain optically thick clumps.

The wind mass-loss rate is largely unaffected by SRSs. This is because
the mass-loss rate is set very close to the star and, as is evident
from Fig.~\ref{fig:general_plots}, the ionization parameter is low
enough in this region to have little affect on the line force. It is
interesting to note that the influence of SRSs inhibits the wind
acceleration which causes the inner wind density to increase, thus
reducing the influence of X-ray ionization ($\xi \propto \rho^{-1}$).

\begin{figure}
  \begin{center}
    \begin{tabular}{c}
\resizebox{80mm}{!}{\includegraphics{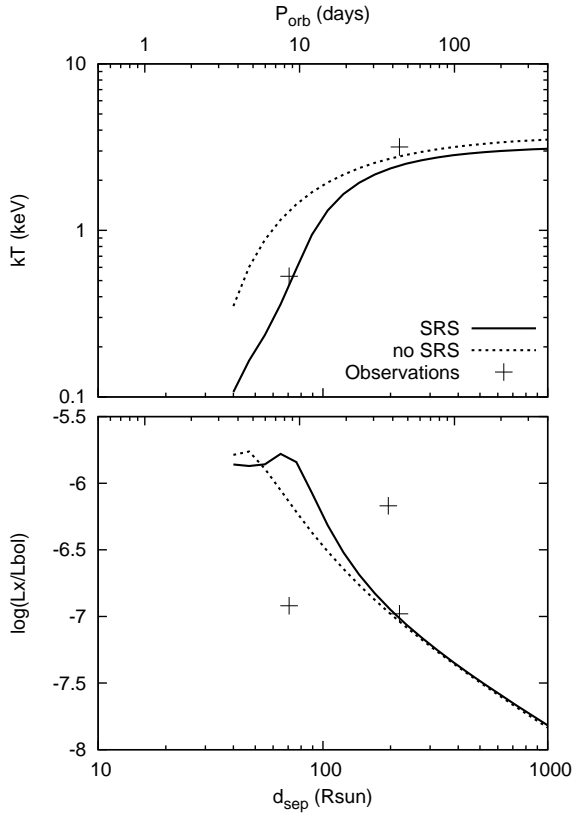}} \\
 \end{tabular}
 \caption{Plots of $kT$ (upper) and $\log(L_{\rm X}/L_{\rm bol})$
   (lower) against binary separation and orbital period (assuming
   circular orbits) for the O6V+O6V binary system. Details of the
   observations are given in \S~\ref{sec:obs_compare}.}
    \label{fig:general_plots3}
  \end{center}
\end{figure}
\subsection{Variation with binary separation}
\label{subsec:binary_sep}

To better understand the region of parameter space in which SRSs will
influence the dynamics of the flow and the observable properties of a
binary system, we have performed further model calculations for the
O6V+O6V binary at a range of binary separations, the results of which
are shown in Figs.~\ref{fig:general_plots2} and
\ref{fig:general_plots3}. As anticipated from \S~\ref{subsec:general},
and illustrated by the calculations with and without SRS, the
mass-loss rate does not vary greatly due to SRSs. The decrease in
mass-loss rate with decreasing binary separation occurs due to the
inhibition of wind acceleration by the opposing star's radiation field
\citep{Stevens:1994}\footnote{\cite{Stevens:1994} note that for a
  ratio of Eddington factors, $\Gamma_{1}/\Gamma_{2} \ltsimm 4$ the
  mass-loss rate will decrease rather than increase compared to the
  single-star case. Furthermore, from their equation (24), it is clear
  that as $d$ decreases, $A_{\rm c}$ increases, causing a reduction in
  $\dot{M}_{\Omega}$}.

Also evident from Fig.~\ref{fig:general_plots2} is that SRSs reduce
the pre-shock wind velocity for a large range of binary separations -
for the O6V+O6V binary the relative difference in $v_{\rm sh}$ between
calculations with and without SRSs is 45$\%$ at $d_{\rm sep}=40\Rsol$
and steadily decreases to 6$\%$ at $d_{\rm sep}=1000 \Rsol$. SRSs also
cause the downturn in $v_{\rm sh}$ to occur at larger separations than
without SRSs.

A secondary effect of a lower $v_{\rm sh}$ is that the importance of
radiative cooling increases (lower $\chi$). As
Fig.~\ref{fig:general_plots2} illustrates, without SRSs we would not
expect radiative shocks ($\chi \ltsimm 1$) until $d_{\rm sep} \ltsimm
40\Rsol$ ($P_{\rm orb}\ltsimm 4\;$days). In contrast, with SRSs this
range increases out to $d_{\rm sep} \ltsimm 70 \Rsol$ ($P_{\rm
  orb}\ltsimm10\;$days). Therefore, we expect that systems will have
radiative shocks for larger separations than previously anticipated.

The inclusion of SRSs causes a reduction in plasma temperature, $kT$,
at all separations but only significantly affects the X-ray
luminosity, $L_{\rm X}$, for a limited range of separations
(Fig.~\ref{fig:general_plots3}). At large separations ($d_{\rm sep}
\gtsimm 200\Rsol$) we do not predict a considerable difference in
$L_{\rm X}$ due to SRSs. Within our model, this stems from the scaling
$L_{\rm X} \propto (\dot{M}/v_{\rm sh})^2$ for adiabatic shocks (see
Eqs~(\ref{eqn:lx}) and (\ref{eqn:chi})), and the differences in
$\dot{M}$ and $v_{\rm sh}$ between models with and without SRSs
(Fig.~\ref{fig:general_plots2}). These competing effects effectively
cancel to produce almost identical X-ray luminosities. For example,
for separations greater than 200\Rsol there are uniform offsets of
roughly $7\%$ for $\dot{M}$ and $v_{\rm sh}$ between calculations with
or without SRSs.

There are, however, a range of separations where SRSs are predicted to
make the system intrinsically brighter. For the O6V+O6V binary this
range is $70 < d_{\rm sep} <200\Rsol$, and arises because $v_{\rm sh}$
decreases more rapidly with decreasing binary separation with SRSs
than without (and because $\Lx \propto (\dot{M}/v_{\rm sh})^2$ at the
relevant values of $\chi$). With SRSs, and at $d_{\rm sep}<70\Rsol$,
$\chi < 1$ therefore $L_{\rm X} \propto \dot{M} v_{\rm sh}^2$ and the
decrease in $v_{\rm sh}$ caused by SRSs reduces the intrinsic
brightness of the wind-wind collision shocks. The abrupt flattening of
$L_{\rm X}$ as the separation is reduced below $d \sim 70\Rsol$ in the
model with SRSs (Fig.~\ref{fig:general_plots3}) is a consequence of
the transition from $\chi > 1$ to $\chi < 1$, which alters the
dependence of \Lx\ on $v_{\rm sh}$.

\begin{figure*}
  \begin{center}
    \begin{tabular}{cc}
\resizebox{80mm}{!}{\includegraphics{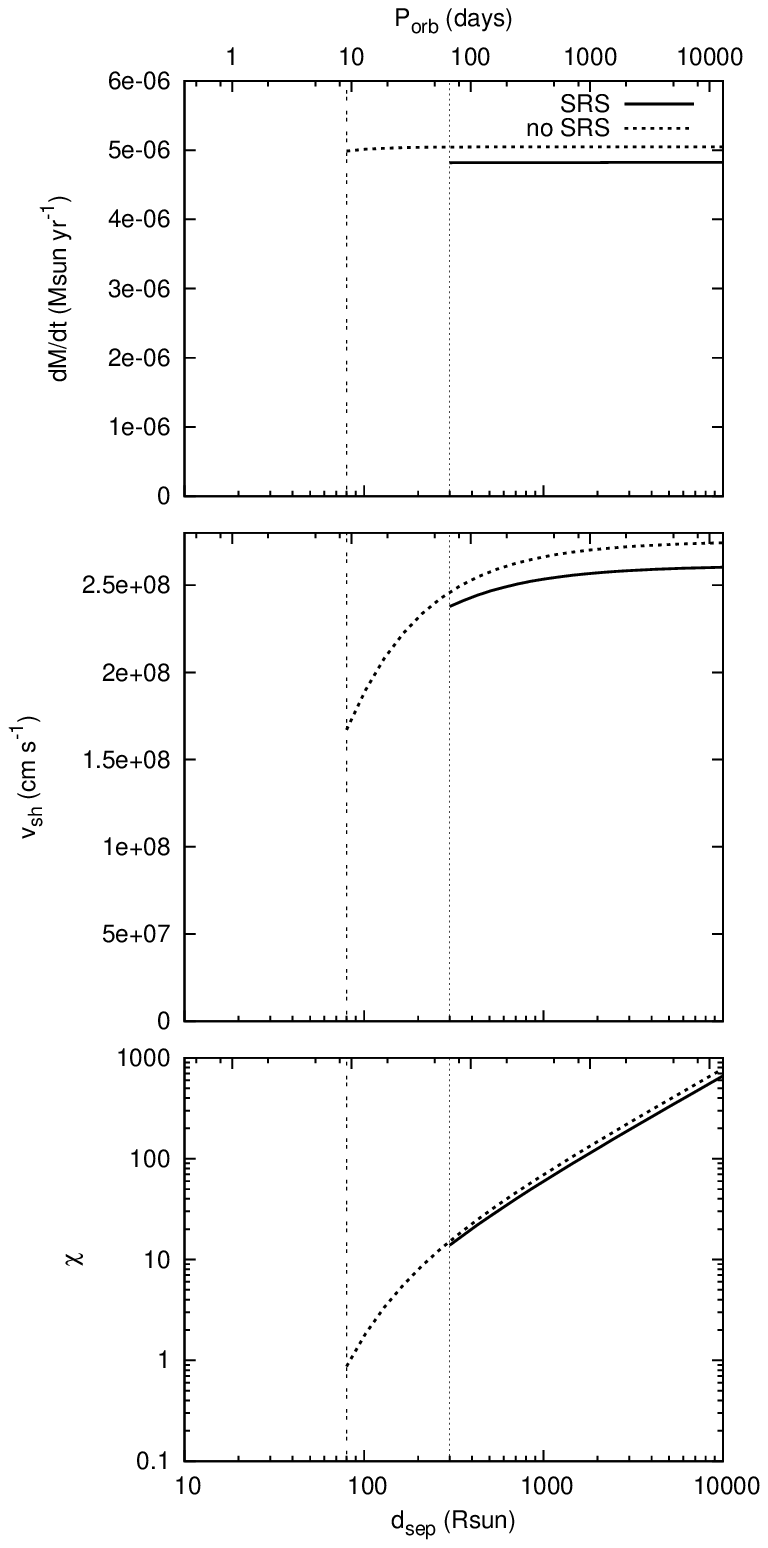}} &
\resizebox{80mm}{!}{\includegraphics{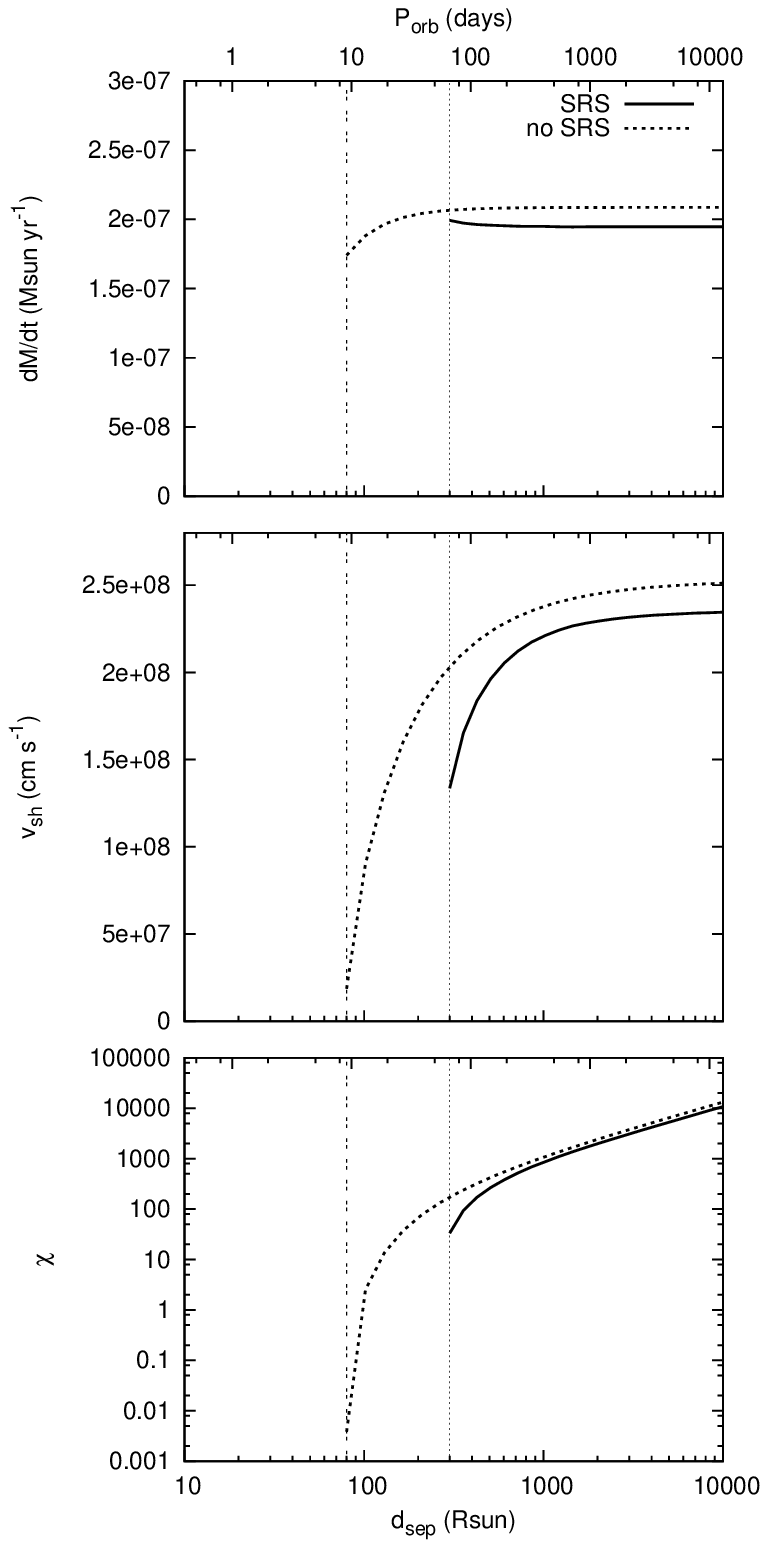}} \\
 \end{tabular}
 \caption{Comparison of calculations for the O4III+O6V binary with and
   without self regulating shocks. Orbital periods are calculated
   assuming circular orbits. Plots are shown for the O4III's wind
   (left column) and the O6V's wind (right column). From top to
   bottom: $\dot{M}$, $v_{\rm sh}$, and $\chi$. The vertical lines
   indicate the limiting separation below which a model solution could
   not be attained for cases with (dotted line) and without SRSs
   (dashed line). At smaller separations, a wind-photosphere
   collision is expected (see \S~\ref{subsec:windphot}).}
    \label{fig:o4iii_o6v}
  \end{center}
\end{figure*}

\begin{figure}
  \begin{center}
    \begin{tabular}{c}
\resizebox{80mm}{!}{\includegraphics{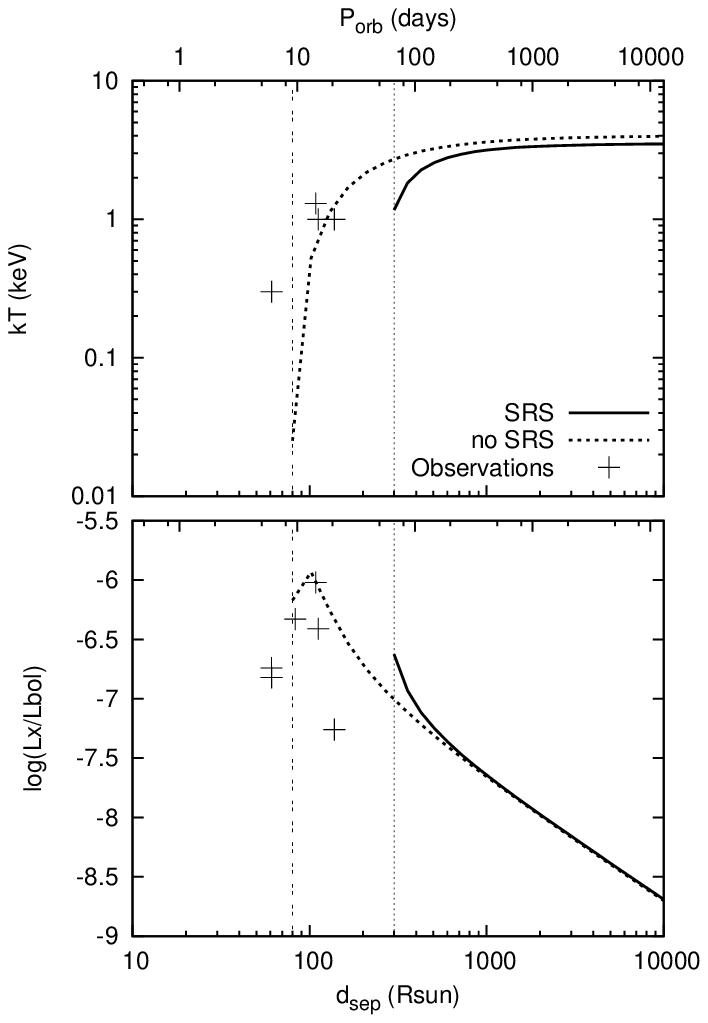}} \\
 \end{tabular}
 \caption{Plots of $kT$ (upper) and $\log(L_{\rm X}/L_{\rm bol})$
   (lower) against binary separation and orbital period (assuming
   circular orbits) for the O4III+O6V binary system. The vertical
   lines indicate the limiting separation below which a model solution
   could not be attained for cases with (dotted line) and without SRSs
   (dashed line). Details of the observations are given in
   \S~\ref{sec:obs_compare}.}
    \label{fig:o4iii_o6v2}
  \end{center}
\end{figure}

\subsection{O4III + O6V binary}

We now consider an O4III+O6V binary with unequal wind momenta. Based
on values for $\dot{M}$ and $v_{\infty}$ from
Table~\ref{tab:stellar_models} (which are calculated using isolated
single star wind models) the wind-wind momentum ratio, $\zeta = 0.04$
(in favour of the O4III star). Assuming terminal velocity winds
(i.e. neglecting radiative inhibition and SRSs), the distance of the
wind-wind momentum balance point from the star with the weaker wind
is,
\begin{equation}
  r_{2} = d_{\rm sep} \frac{1+\zeta^{1/2}}{\zeta^{1/2}}.
\end{equation}
Setting $r_{2}=R_{\ast 2}=10.2\Rsol$, one estimates the wind-wind
collision to remain away from the surface of the O6V star for
separations greater than 61\Rsol. If we improve on this estimate using
our wind model {\it without} the inclusion of SRSs we find a stable
wind-wind collision down to separations of 80\Rsol (dashed vertical
lines in Figs.~\ref{fig:o4iii_o6v} and \ref{fig:o4iii_o6v2}). However,
when SRSs are included a stable wind-wind collision is not predicted
to occur for separations smaller than 300\Rsol (dotted vertical lines
in Figs.~\ref{fig:o4iii_o6v} and \ref{fig:o4iii_o6v2}). The reason for
this drastic increase is that SRSs tend to make the weaker wind even
weaker as it is closer to the source of the X-rays at the wind-wind
collision. The tendency for SRSs to reduce the strength of the weaker
wind, therefore, becomes more pronounced as the separation of the
stars is reduced. Consequently, even for comparatively large
separations, wind-launching fails. This general result states that
SRSs will cause a wind-photosphere collision in unequal winds systems
up to larger separations than otherwise expected.

For sufficiently large separations ($d_{\rm sep} > 300\Rsol$) a
wind-wind collision is predicted to occur, and in this regime SRSs
introduce a minor reduction in $\dot{M}$'s of roughly $4\%$ for both
stars (Fig.~\ref{fig:o4iii_o6v}). SRSs also cause a reduction in
$v_{\rm sh}$ for both winds, most notably at smaller separations,
where a sharp downturn arises in $v_{\rm sh}$ for the O6V wind,
signifying the sudden failing of the wind-wind collision as the O6V's
wind becomes increasingly weakened by the X-rays from the
shocks. Somewhat surprisingly, although we anticipate that a
wind-photosphere collision will ensue for relatively large
separations, the post-shock gas is expected to be adiabatic ($\chi \gg
1$). This differs from previous models in which a wind-photosphere
collision is typically accompanied by highly radiative shocks from the
weaker wind \citep{Pittard:1998, Parkin_Gosset:2011}.

Similar to the O6V+O6V binary, $L_{\rm X}$ is largely unaffected by
SRSs for the O4III+O6V binary (Fig.~\ref{fig:o4iii_o6v2}). A
noticeable reduction in $kT$ values is, however, introduced
particularly for smaller separations. In this case we find that SRSs
introduce an offset in $kT$ values for separations larger than
700\Rsol ($P_{\rm orb}>300\;$days), and that for closer separations
SRSs cause a sharp downturn in $kT$, reflecting the behaviour of
$v_{\rm sh}$ for the O6V's wind -
Fig.~\ref{fig:o4iii_o6v}. \Lx~remains similar between the models, with
SRSs causing an upturn in \Lx~for $d_{\rm sep} \simeq 300-600\Rsol$.

We note that although the O4III star has the stronger wind, the X-ray
emission is greater from the shocked O6V's wind because a larger
fraction of its wind is shocked at an angle close to the shock normal
\citep[thus converting a larger fraction of its kinetic energy into
thermal energy -][]{Pittard_Stevens:2002}. It follows that the
observed $kT$ will also be predominantly weighted by the weaker
wind. For the specific parameters used in our model, and at
separations less than 1000 \Rsol ($P_{\rm orb}<400\;$days) the ratio
of X-ray luminosity from the winds is 2:1 in favour of the O6V.

\subsection{O4III + O6V binary with a wind photosphere collision}
\label{subsec:windphot}

\begin{figure}
  \begin{center}
    \begin{tabular}{c}
\resizebox{80mm}{!}{\includegraphics{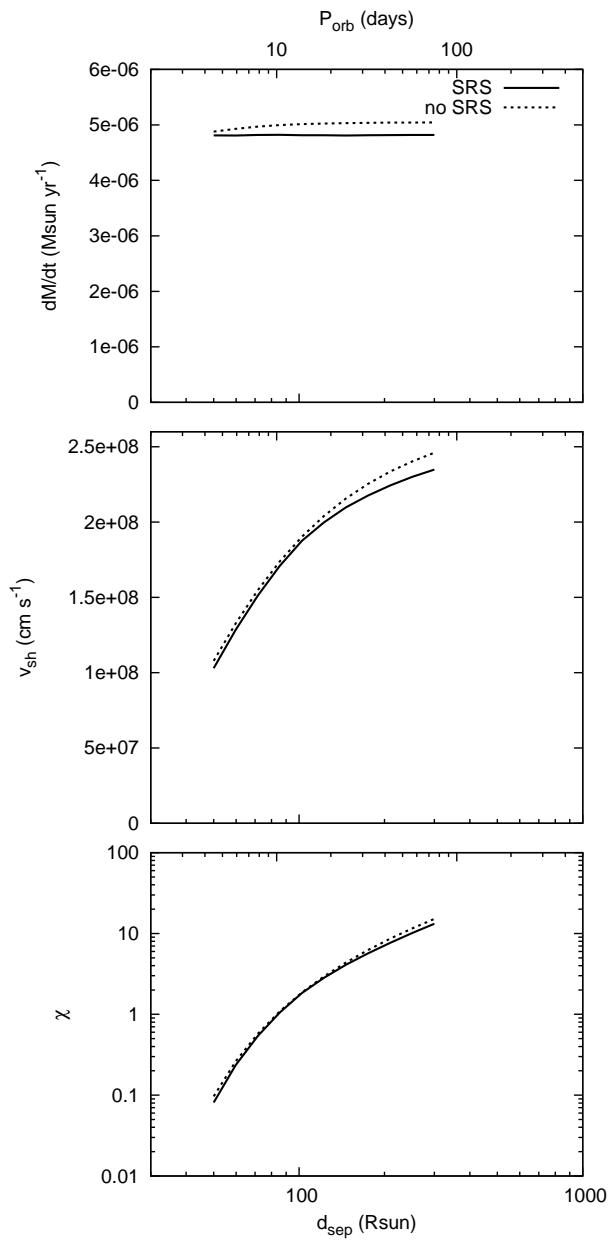}} \\
 \end{tabular}
 \caption{Comparison of calculations for the O4III+O6V binary with a
   wind-photosphere collision and with and without self regulating
   shocks. Orbital periods are calculated assuming circular
   orbits. From top to bottom: $\dot{M}$, $v_{\rm sh}$, and
   $\chi$. (Results are only shown for the O4III's wind because the
   O6V's wind is assumed to be suppressed.)}
    \label{fig:o4iii_phot}
  \end{center}
\end{figure}

\begin{figure}
  \begin{center}
    \begin{tabular}{c}
\resizebox{80mm}{!}{\includegraphics{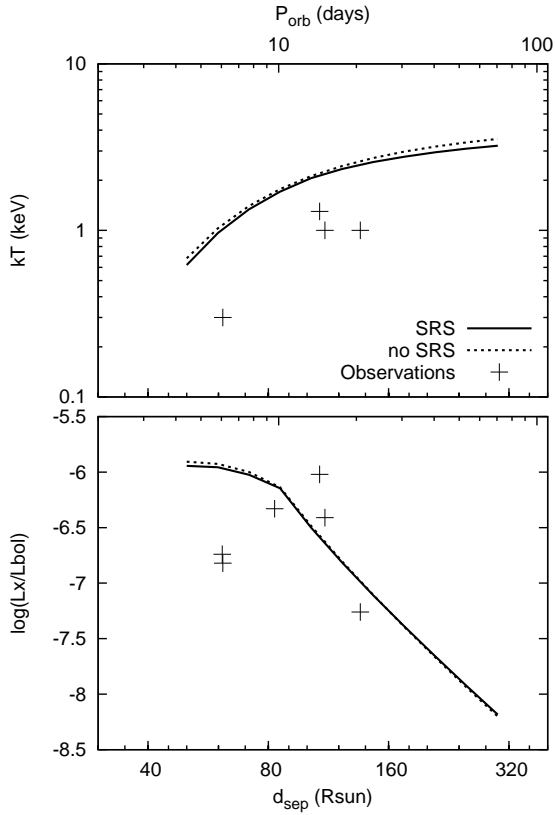}} \\
 \end{tabular}
 \caption{Plots of $kT$ (upper) and $\log(L_{\rm X}/L_{\rm bol})$
   (lower) against binary separation and orbital period (assuming
   circular orbits) for the O4III+O6V binary with a wind-photosphere
   collision. Details of the observations are given in
   \S~\ref{sec:obs_compare}.}
    \label{fig:o4iii_phot2}
  \end{center}
\end{figure}

As mentioned in the preceding section, the inclusion of SRSs
\citep[and radiative inhibition -][]{Stevens:1994} considerably
increases the range of binary separations where a wind-photosphere
collision will occur in a massive star binary system. For our
O4III+O6V model we found that a ram pressure balance between the winds
could not be achieved for separations less than 300\Rsol. In this
section we consider the wind-photosphere collision occurring at
$d_{\rm sep}<300\Rsol$. For this purpose we use the model described in
\S~\ref{sec:model} with the difference that the O6V's wind is not
included and the shock is assumed to occur at the surface of the
companion star, $r=d_{\rm sep}-R_{\ast 2}$. The fractional wind
kinetic power that is thermalized, $\Xi$ is approximated by the solid
angle subtended by the O6V star as viewed by the O4III star (see
\S~\ref{subsec:shocks}).

\begin{figure}
  \begin{center}
    \begin{tabular}{c}
\resizebox{80mm}{!}{\includegraphics{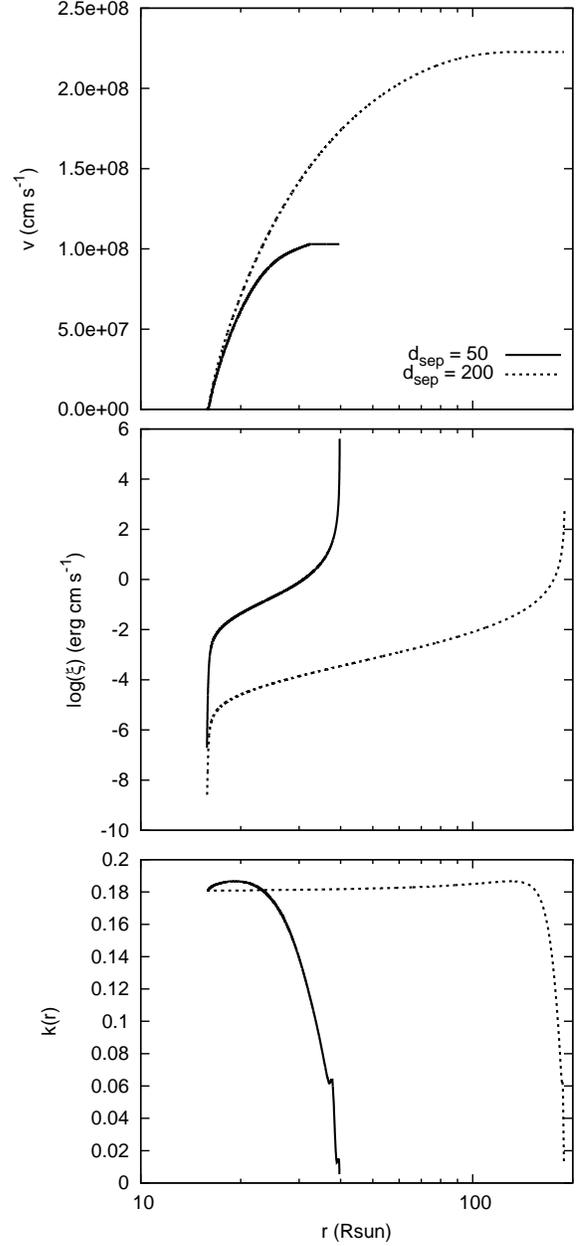}} \\
\end{tabular}
\caption{Plots of $v$ (upper), $\log(\xi)$ (middle), and $k(\xi)$
  (lower) as a function of radius for the O4III+O6V wind-photosphere
  collision model. Curves are shown for calculations at $d_{\rm
    sep}=$50 and 200 \Rsol.}
    \label{fig:o4iii_phot_rad}
  \end{center}
\end{figure}

As is clear from the plots of $\dot{M}$, $v_{\rm sh}$, and $\chi$ in
Fig.~\ref{fig:o4iii_phot} and $kT$ and $\log(L_{\rm X}/L_{\rm bol})$
in Fig.~\ref{fig:o4iii_phot2}, SRSs have very little affect on the
wind-photosphere collision. This is because $\log(\xi) <0$ in the
inner wind acceleration region which allows the wind to accelerate to
a similar velocity to the case with no SRSs. $\log(\xi)>0$ is only
reached in regions beyond the acceleration zone, meaning that the
driving is unaffected. To illustrate this, in
Fig.~\ref{fig:o4iii_phot_rad} we show radial profiles of wind
velocity, $\log(\xi)$, and $k(\xi)$ computed for binary separations of
50 and 200 \Rsol. Clearly, the line force is only suppressed by SRSs
in regions where $\log(\xi) > 0$, which reflects the almost step
function like behaviour of $k(\xi)$ at $\log(\xi) \simeq 0-1$
(Fig.~\ref{fig:k_etamax}). 

\section{An approximate indicator for self regulating shocks}
\label{sec:approx}

It would be useful to have a simple means of estimating the separation
at which we expect SRSs to play an important role in the wind-wind
collision. This could be used, for example, to estimate whether SRSs
should be considered when modelling a specific system. Surveying the
results of our model calculations, we find that, to within an accuracy
of a factor of two, we have the following approximate relations for
our equal winds binary system:
\begin{equation}
  v_{\rm sh} \approx 2 v_{\rm esc}, \label{eqn:vsh_approx}
\end{equation}
\begin{equation}
  v_{\rm esc} \approx \sqrt{2 G M_{\ast}/R_{\ast}},
\end{equation}
\begin{equation}
  L_{\rm X} = \frac{\Xi}{1 + \chi} \dot{M} v_{\rm sh}^2\approx
  \frac{d_{0}}{d_{\rm sep}} \dot{M} v_{\rm sh}^2,
\end{equation}
\begin{equation}
  \rho(2 R_{\ast}) \approx \frac{\dot{M}}{16 \pi v_{\rm
      esc}R_{\ast}^2}, \label{eqn:rho_approx}
\end{equation}
where $\rho(2 R_{\ast})$ is the wind density at a radius of
$2~R_{\ast}$ and $d_{0}=0.5 \Rsol$ is a constant used to fit the
variation of $\Xi/(1 + \chi)$ with $d$. With the further
simplification of neglecting any attenuation of X-rays as they travel
back through the wind towards the star (which has been shown in
\S~\ref{subsec:general} to have minor influence, at least for the
system parameters considered), inserting
Eqs~(\ref{eqn:vsh_approx})-(\ref{eqn:rho_approx}) into
Eq~(\ref{eqn:xi}) gives,
\begin{equation}
  \xi(2 R_{\ast}) = 64 \pi^2 \mu m_{\rm H} d_{0}\frac{v_{\rm
      esc}^3}{d_{\rm sep}}\left(\frac{d_{\rm sep}}{4 R_{\ast}} - 1\right)^{-2}. \label{eqn:xi_approx}
\end{equation}
From the results of \S~\ref{sec:results} a requirement for SRSs to
affect the wind-wind collision is that $\log(\xi(2 R_{\ast})) \gtsimm
0$. Setting $\xi(2 R_{\ast})=1$ and re-arranging
Eq~(\ref{eqn:xi_approx}) leads to a cubic equation for $d_{\rm sep}$
which, for parameters pertaining to our equal winds O6V+O6V binary
(\S\S~\ref{subsec:general} and \ref{subsec:binary_sep}), has one real
root of $d_{\rm sep} \simeq 91\Rsol$. Inspecting
Figs.~\ref{fig:general_plots2} and \ref{fig:general_plots3} one sees
that this value is consistent with the onset of a marked difference
due to SRSs.

\section{Comparison against observations}
\label{sec:obs_compare}

To facilitate a comparison of our model results against observed O+O
binaries we have extracted a sample of systems from the studies by
\cite{Gagne:2011} and \cite{Gagne:2012}. Only systems with orbital
periods within the range of our models have been considered. We then
separated the remaining systems into those with roughly equal winds or
unequal winds systems, where we classify the former as systems in
which the stars differ by less than a spectral type and/or subclass,
and the latter as systems which differ by more than this
increment. The roughly equal winds systems (three in total) are then
compared to our O6V+O6V binary and the unequal winds systems (six in
total - although $kT$ values are only available for four systems)
against the O4III+O6V binary.

We remind the reader that the model used in the current investigation
includes simplifications to the physics, which have been chosen so as
to allow a tractable initial exploration of the SRS effect and its
potential importance. Nevertheless, in this section we compare our
model results against observations to provide a sense of how SRSs
might be relevant in explaining general trends. However, we caution
that detailed comparison must await more thorough modelling.

\subsection{Equal winds systems}

In the equal winds case we find reasonably good agreement between our
O6V+O6V model and the observed $kT$ and $\log(L_{\rm X}/L_{\rm bol})$
(Fig.~\ref{fig:general_plots3}). Including SRSs improves the match to
the observed $kT$ values. The O6V+O6V results
(Fig.~\ref{fig:general_plots3}) show that for binary separations less
than $\sim200\Rsol$ ($P_{\rm orb}<30\;$days) we expect roughly equal
winds systems to be brighter than the expected luminosity from
embedded wind shocks in the respective stars \citep[$\log(L_{\rm
  X}/L_{\rm bol}) \simeq -7$ , e.g.,][]{Sana:2006, Naze:2011}. Roughly
equal winds systems with separations larger than $\sim200\Rsol$ will
not, therefore, be identifiable as CWB systems from their X-ray
luminosity but instead they may be identifiable by $kT> 0.6\;{\rm
  keV}$ \citep[i.e. hotter plasma temperature than anticipated from a
single massive star -][]{Owocki:1999}. Indeed, it may be the case that
only early-type O+O binaries with intermediate orbits and strong winds
will be prolific X-ray emitters \citep[e.g., Cyg OB\#9 - $P_{\rm
  orb}=858\;$days -][]{Naze:2012}, with the majority of later-type
massive binaries only being identifiable as CWBs (in X-rays) via
plasma temperatures above 0.6~keV.

There is significant scatter in the observed $kT$ values and
$\log(L_{\rm X}/L_{\rm bol})$ for orbital periods less than six
days. We do not attempt to compare our model against these systems
because, when the separation of the stars becomes comparable to their
stellar radii, one expects additional effects that we have not
considered to become important. For example, tidal deformation,
gravity darkening, photospheric reflection, and the possibility of
mass transfer \citep{Gayley:1999, Dessart:2003, Owocki:2007, Dermine:2009}.

\subsection{Unequal winds systems}

At the separations of the observed unequal winds binaries, a wind-wind
collision is predicted from models without SRSs while a
wind-photosphere collision is expected based on our SRS
calculations. Comparing Figs.~\ref{fig:o4iii_o6v2} and
\ref{fig:o4iii_phot2} one sees that both cases do arguably similarly
well at matching the observations - although the wind-wind collision
with no SRSs does appear to over-predict the observed $\log(L_{\rm
  X}/L_{\rm bol})$. Interpreting this comparison is complicated, and
it may simply be indicating that reality lies between these two
different cases, i.e. a wind-wind collision prevailing to smaller
separations but with some wind suppression due to SRSs.

Examining the wind-photosphere collision in more detail, the models
systematically over-predict $kT$ values by roughly a factor of two,
irrespective of whether SRSs are considered or not
(Fig.~\ref{fig:o4iii_phot2}). However, the agreement between the model
and observations is reasonably good for $\log(L_{\rm X}/L_{\rm bol})$,
with the exception of the two systems with orbital periods of roughly
6 days: HD93205 \citep{Townsley:2011, Naze:2011} and HD101190
\citep{Chlebowski:1989, Sana:2011, Gagne:2012}. We remind the reader
that orbital periods have been converted to binary separations under
the basic assumption of circular orbits, which is accurate for the
majority of the systems in the sample. Considering the two outliers
with orbital periods of roughly 6 days, the former, HD93205, has an
orbital eccentricity of 0.37 \citep{Morrell:2001, Rauw:2009}. Using
the ephemeris from \cite{Morrell:2001} and the date of the {\it
  Chandra} observation of HD93205, we estimate an orbital phase of
$\sim 0.2$. As this is relatively close to periastron, we cannot
appeal to the larger separation that will occur at apastron to improve
the match against our model results. Adopting the recently derived
orbital solution for HD101190 with an eccentricity of $\sim0.3$
\citep{Sana:2011} does not help the agreement between our model and
its $\log(\Lx/L_{\rm bol})$ datapoint either. A more detailed
hydrodynamical model of a wind-photosphere collision is warranted to
investigate the systematic discrepancy in $kT$ values and
$\log(\Lx/L_{\rm bol})$.

\section{Discussion}
\label{sec:discussion}

\cite{Owocki:1995} and \cite{Gayley:1997} have argued that for binary
systems where a ram pressure balance is not expected to occur, the
radiation field of the star with the weaker wind may decelerate the
incoming wind of its stronger companion. We note that the high values
of $\xi$ in the vicinity of the shocks (see
Figs.~\ref{fig:general_plots} and \ref{fig:o4iii_phot_rad}) raises
questions about the ability of radiative braking to produce a
time-steady wind interaction region. For instance, perhaps an incoming
flow is initially subject to radiative braking, but any shock which
subsequently forms (and the associated X-ray flux) will suppress the
braking force, leading to the dominant wind continuing on its path
towards the weaker star's photosphere. Then, with the X-ray emitting
shocks extinguished - or sufficiently weaker/far enough away from the
point where radiative braking was originally effective - the cycle can
repeat. However, more detailed hydrodynamical model is required to
properly assess these points as it may be the case that close to the
photosphere of a star the gas density will be sufficiently high that
the ionization parameter will be small (either due to an intrinsically
dense photosphere, a build-up of gas behind the shock, or wind
strengths weakened by inhibition/braking) , in which case radiative
braking may prevail. Therefore, a complicated, and most likely
time-dependent, competition between radiative braking and SRSs may
arise.

In our model calculations we have examined the {\it intrinsic} X-ray
luminosity. An important related question is how the {\it observed}
$L_{\rm X}$ would be affected by shock self-regulation? For instance,
SRSs reduce the post-shock gas temperature and, consequently, the
energy of emitted X-rays. As the susceptibility of X-rays to
absorption increases at lower energies, the X-ray flux that reaches
the observer may be comparatively much fainter for systems where SRSs
are effective. Therefore, although our model including SRSs
overestimates the observed \Lx~for binary systems with orbital periods
of a few days to about the same level as model cwb1 from
\cite{Pittard:2009} \citep[see also][]{Pittard_Parkin:2010}, further
work is needed to evaluate how the observed (attenuated) \Lx~is
impacted by SRSs. This will be an important point to pursue in future
work.

The model adopted for this investigation features a number of
approximate relations, adopted to keep the calculations relatively
simple whilst achieving an order-of-unity accurate prediction of the
influence of SRSs on a wind-wind collision. While these approximations
have been chosen in order to give a simple exposition of the SRS
mechanism, it is important to note that alternative approximations
could have been made, whose respective merits should be borne in mind
for future investigations. Firstly, the approximation used to estimate
the half-opening angle of the bow shock, $\theta_{\rm half}=(\pi
\zeta_{\rm eff})/(1+ \zeta_{\rm eff})$, which features in
Eq~(\ref{eqn:zabalza_xi}) for $\Xi$ does not consider the influence of
the post-shock wind momentum on the global shock
geometry. Calculations of shock half-opening angles which include this
additional momentum flux \citep[e.g.][]{Canto:1996, Gayley:2009} find
that it widens the bow shock, leading to a slightly different, and
more accurate, scaling of $\theta_{\rm half}$ with $\zeta_{\rm
  eff}$. Similarly, a more accurate expression for $\Xi$, in the case
of a wind-photosphere collision (Eq~\ref{eqn:xi_wind_phot}) could
likely be derived using a global momentum flux approach similar to
that adopted by \cite{Gayley:2009}.  Secondly, in using
Eq~(\ref{eqn:zabalza_xi}) to calculate the thermalization efficiency
it is implicitly assuming that all wind kinetic energy normal to the
shock is thermalized and the effect of shock obliquity is not included
(although a constant obliquity correction is included when calculating
the mean plasma temperature). We anticipate that a more accurate
treatment of shock obliquity would introduce an order unity correction
to the results and could improve the agreement with
observations. Thirdly, we do not include the radiative-driving force
arising from the stellar radiation field reflected by the opposing
star's photosphere. \cite{Gayley:1999} examined a similar scenario in
planar geometry and found that the radiative inhibition effect
\citep{Stevens:1994} was weaker, and the mass-loss rate higher, due to
the extra acceleration force from reflected radiation. To include the
reflection effect in a geometry such as illustrated in
Fig.~\ref{fig:srs_cartoon} is not trivial, but would be a worthwhile
avenue for future work. Reflection could be particularly important for
the wind-photosphere collision model as it could enhance radiative
braking and/or impinge on the wind-bearing star's wind acceleration.
\section{Conclusions}
\label{sec:conclusions}

We have presented steady-state wind models for massive star binary
systems in which the X-ray emission from the wind-wind collision
shocks modifies the driving of the wind, which we term self regulating
shocks (SRSs). To this end we include a parameterized radiative line
force with X-ray ionization dependence (derived from line force
calculations) in our wind model. Our primary result is that X-ray
radiation from the shocks is found to inhibit the wind acceleration
and can lead to lower pre-shock velocities, which in turn causes the
post-shock plasma temperature to decrease. In general, SRSs will alter
the pre-shock velocity if the ionization parameter $\log(\xi) \gtsimm
0$ within a radius of $2~R_{\ast}$. We believe the qualitative results
from this investigation to be robust, but note that quantitative
estimates made from the model may change as more complete physics
prescriptions are incorporated into new models. Caution should be
exercised when extrapolating the results from the sample calculations
in this paper to specific systems, as to acquire accurate results will
require a dedicated analysis.

Despite the presence of an anticipated feedback loop between the
shocks and the wind driving, the resulting {\it intrinsic} X-ray
luminosity of the shocks is not strongly altered by the inclusion of
SRSs. However, although not examined in this work, lower plasma
temperatures may render the X-ray emission more susceptible to
absorption, which could have an impact on the observed {\it
  attenuated} emission.

We have presented model results for O6V+O6V and O4III+O6V binary
systems computed for a wide range of binary separations. For the
O6V+O6V binary the main difference introduced by SRSs is the reduction
in pre-shock velocities described above. For the O4III+O6V binary,
SRSs greatly increase the separation at which a wind-photosphere
collision (which occurs when there is no ram pressure balance between
the winds) from 80 to 300 \Rsol. Furthermore, close to the shocks,
where X-ray ionization is greatest, the line force can be completely
suppressed, and we conjecture that this may render radiative braking
ineffective, or highly time-dependent.

A comparison of our model results to observations reveals that the
inclusion of self-regulated shocks improves the agreement for plasma
temperatures in roughly equal winds systems. However, irrespective of
the inclusion of self-regulated shocks we find a systematic offset in
plasma temperatures for unequal winds systems (which we model as a
wind-photosphere collision, as expected for the range of binary
separations probed by observations). The models show reasonable
agreement with observations for $\log(L_{\rm X}/L_{\rm bol})$. Unequal
winds O+O star systems with a wind-wind collision are not expected to
be brighter than their respective stars in X-rays. Such systems are
predicted only to have a wind-wind collision above some cutoff binary
separation because at smaller separations SRSs prevent a stable
wind-wind ram pressure balance. For our sample O4III+O6V system, this
cutoff is at a separation of $300\Rsol$ ($P_{\rm
  orb}>70\;$days). However, shorter period systems (separations
smaller than $130\Rsol$, $P_{\rm orb}<20\;$days for our O4III+O6V
model) with a wind-photosphere collision should be noticeably bright
in X-rays (i.e. $\log(L_{\rm X}/L_{\rm bol}) > -7$).

This work is a first attempt at modelling the influence of X-ray
ionization on wind driving in massive star binary systems. In closing
we suggest a few possible avenues for future work. Developing more
realistic models requires multi-dimensionality, with 2D models being
the logical next step. Furthermore, time-dependent calculations would
be enlightening as one can envisage that oscillatory behaviour may
result from perturbations in the pre-/post-shock flow, and it will be
interesting to examine whether SRSs can explain flaring in massive
star binary X-ray lightcurves \citep[e.g.][]{Moffat:2009}.

\subsection*{Acknowledgements}
We gratefully thank the referee, Ken Gayley, for a useful and
informative report that helped to improve the paper, and Julian
Pittard for helpful comments on an earlier draft. E.~R.~P thanks the
Australian Research Council for funding through the Discovery Projects
funding scheme (project number DP1096417).


\appendix
\section{Tables of fits to line force parameters}
\label{sec:appendix}

In \S~\ref{sec:line_force} we described the dependence of the line
force on the ionization parameter, $\xi$. To allow a straightforward
application to the wind model in \S~\ref{sec:model} we described this
dependence in terms of the parameters $k(\xi)$ and $\eta_{\rm
  max}(\xi)$. Tabulated values of these parameters are provided in
Table~\ref{tab:fits}. Note that these values have not been rescaled
(as described in \S~\ref{subsec:rescale}). For $\log(\xi)>4$ the line
force is strongly suppressed ($M \sim 0$).

\begin{table}
\begin{center}
  \caption[]{Fit parameters for $k(\xi)$ and $\eta_{\rm max}(\xi)$} \label{tab:fits}
\begin{tabular}{lllll}
\hline
$\log(\xi)$ & \multicolumn{2}{c}{O6V}  &  \multicolumn{2}{c}{O4III} \\
 & $k$ & $\log(\eta_{\rm max})$ & $k$ & $\log(\eta_{\rm max})$ \\
\hline
-5.0&  0.348&  7.035&  0.244&  6.705\\
 -4.0&  0.349&  7.034&  0.245&  6.703\\
 -3.0&  0.351&  7.029&  0.247&  6.699\\
 -2.5&  0.353&  7.023&  0.248&  6.693\\
 -2.0&  0.356&  7.015&  0.250&  6.685\\
 -1.5&  0.358&  7.003&  0.252&  6.672\\
 -1.0&  0.351&  6.992&  0.248&  6.662\\
 -0.5&  0.321&  6.992&  0.223&  6.675\\
  0.0&  0.263&  6.999&  0.176&  6.722\\
  0.5&  0.196&  6.966&  0.128&  6.721\\
  1.0&  0.130&  6.526&  0.090&  6.288\\
  1.5&  0.103&  4.891&  0.087&  4.268\\
  2.0&  0.035&  4.164&  0.029&  3.392\\
  2.5&  0.021&  2.660&  0.018&  2.305\\
  3.0&  0.023&  2.068&  0.020&  1.911\\
  3.5&  0.020&  1.655&  0.019&  1.563\\
  4.0&  0.014&  1.233&  0.015&  1.220\\
\hline
\end{tabular}
\tablecomments{The \cite{Castor:1975} parameter $\alpha$ is fixed for
  each star, respectively (see Table~\ref{tab:stellar_models} and
  \S~\ref{sec:line_force}).}
\end{center}
\end{table}

\label{lastpage}

\end{document}